\documentclass[fleqn,usenatbib]{mnras}

\usepackage{mathptmx}

\usepackage[T1]{fontenc}

\usepackage{amssymb}	
\usepackage{graphicx}	
\usepackage{amsmath}	

\usepackage{hyperref}
\hypersetup{colorlinks=true,linkcolor=blue,citecolor=blue,filecolor=blue,urlcolor=blue}

\usepackage{siunitx}    

\usepackage[usenames, dvipsnames]{color}
\usepackage{soul}

\usepackage{subcaption}

\title[Radio-AGN across the galaxy population]{Radio-AGN activity across the galaxy population: dependence on stellar mass, star-formation rate, and redshift}

\author[R. Kondapally et al.]{Rohit~Kondapally,$^{1}$\thanks{E-mail: rohitk@roe.ac.uk}
Philip~N.~Best,$^{1}$
Kenneth~J.~Duncan,$^{1}$
Huub~J.~A.~R\"{o}ttgering,$^{2}$
Daniel~J.~B.~Smith,$^{3}$
\newauthor{Isabella~Prandoni,$^{4}$
Martin~J.~Hardcastle,$^{3}$
Tanja~Holc,$^{1}$
Abigail~L.~Patrick,$^{1}$
Marina~I.~Arnaudova,$^{3}$}
\newauthor{Beatriz~Mingo,$^{3,5}$
Rachel~K.~Cochrane,$^{1,6}$
Soumyadeep~Das,$^{3}$
Paul~Haskell,$^{3}$
Manuela~Magliocchetti,$^{7}$}
\newauthor{Katarzyna~Ma{\l}ek,$^{8}$
George~K.~Miley,$^{2}$
Cyril~Tasse$^{9,10}$
and Wendy~L.~Williams,$^{11}$
}
\\
$^{1}$Institute for Astronomy, University of Edinburgh, Royal Observatory, Blackford Hill, Edinburgh, EH9 3HJ, UK \\
$^{2}$Leiden Observatory, Leiden University, PO Box 9513, NL-2300 RA Leiden, the Netherlands\\
$^{3}$Centre for Astrophysics Research, University of Hertfordshire, College Lane, Hatfield AL10 9AB, UK\\
$^{4}$INAF-Istituto di Radioastronomia, Via Gobetti 101, I-40129, Bologna, Italy\\
$^{5}$School of Physical Sciences, The Open University, Walton Hall, Milton Keynes, MK7 6AA, UK\\
$^{6}$Department of Astronomy, Columbia University, New York, NY 10027, USA\\
$^{7}$INAF-IAPS, Via Fosso del Cavaliere 100, 00133, Rome, Italy\\
$^{8}$National Centre for Nuclear Research, Pasteura 7, 02-093 Warsaw, Poland\\
$^{9}$GEPI \& ORN, Observatoire de Paris, Université PSL, CNRS, 5 Place Jules Janssen, 92190 Meudon, France\\
$^{10}$Department of Physics \& Electronics, Rhodes University, PO Box 94, Grahamstown, 6140, South Africa\\
$^{11}$SKA Observatory, Jodrell Bank, Lower Withington, Macclesfield, SK11 9FT, United Kingdom\\
}

\date{Accepted XXX. Received YYY; in original form ZZZ}

\pubyear{2024}

\begin{document}
\label{firstpage}
\pagerange{\pageref{firstpage}--\pageref{lastpage}}
\maketitle

\begin{abstract}
We characterise the co-evolution of radio-loud AGN and their galaxies by mapping the dependence of radio-loud AGN activity on stellar mass and star-formation rate (SFR) across cosmic time (out to $z \sim 1.5$). Deep LOFAR radio observations are combined with large galaxy samples to study the incidence of radio-loud AGN across the galaxy population; the AGN are further split into low-excitation radio galaxies (LERGs) and high-excitation radio galaxies (HERGs). We find that LERG activity occurs over a wide range of SFRs, whereas HERGs are typically found in galaxies with ongoing star formation. The LERGs are then split based on their SFRs relative to the main sequence, across redshift. Within quiescent galaxies, LERG activity shows a steep stellar mass dependence with the same normalisation across the past $\sim$ 10\,Gyr; this indicates that hot gas fuels LERGs in quiescent galaxies across cosmic time. In massive galaxies ($\log_{10}(M/\rm{M_{\odot}}) \gtrsim 11$), the incidence of LERGs is roughly constant across the galaxy population, suggesting that LERGs in massive galaxies may be fuelled by hot gas regardless of the star-formation activity. At lower masses, however, LERG activity is significantly more enhanced (by a factor of up to 10) in star-forming galaxies compared to quiescent galaxies; this suggests that an additional fuelling mechanism, likely associated with cold gas, may fuel the LERGs in galaxies with higher SFRs. We find that HERGs typically accrete above 1 per cent of the Eddington-scaled accretion rate, and the LERGs typically accrete below this level.
\end{abstract}




\section{Introduction}\label{sec:intro}
It is now widely believed that supermassive black holes (SMBHs) and their host galaxies co-evolve across cosmic time, with evidence for a link between the build-up of stars and the growth of the SMBHs \citep[e.g.][]{2013ARA&A..51..511K}, and a tight correlation between the mass of the SMBH and that of the galaxy bulge component \citep[e.g.][]{2000ApJ...539L...9F,2001MNRAS.320L..30M}. SMBHs can undergo an active phase, powered by accretion of matter onto the black hole, during which they are referred to as active galactic nuclei (AGN). These AGN can emit vast amounts of energy across the electromagnetic spectrum in the form of ionising outflows or relativistic jets, which can suppress or regulate subsequent star-formation within the host galaxy; this is referred to as AGN feedback \citep[e.g.][]{best2005_sdss_xmatch,cattaneo2009_feedback_review,2012ARA&A..50..455F,2014ARA&A..52..589H,Hardcastle2020}. AGN feedback is often invoked in cosmological simulations to suppress the growth of the most massive haloes and reproduce the observed local galaxy luminosity functions \citep[e.g.][]{2006MNRAS.370..645B,2006MNRAS.365...11C,Somerville2015}.

Of particular importance in the lifecycle of massive galaxies and clusters are radio-loud AGN (hereafter; radio-AGN), which emit powerful bi-polar jets of relativistic ionised material that radiate synchrotron emission, which is visible at radio wavelengths. The energy deposited by these radio-jets into the host galaxy and surrounding environment can balance the radiative cooling losses and regulate star-formation in massive galaxies, with recurrent radio-AGN heating required to `maintain' galaxies as `red and dead' once quenched, in the nearby Universe \citep[see][]{Best2006,2007MNRAS.379..894B,2007ARA&A..45..117M,2014ARA&A..52..589H,Hardcastle2020}.

Radio-AGN are typically classified into two modes, based on the nature of the emission lines in their optical spectra, as low-excitation radio galaxies (LERGs) and high-excitation radio galaxies (HERGs). This classification is understood to be linked to the nature of the accretion flow onto the SMBH \citep[see][]{Best2005fagn,Allen2006,Hardcastle2007,Hardcastle2018}, where HERGs are associated with radiatively-efficient accretion, typically from cold gas, which leads to the formation of a geometrically thin, optically thick accretion disc and a dusty torus \citep[e.g.][]{1973A&A....24..337S}; as a result, they display high-excitation emission lines due to photoionization of gas from accretion disc photons. In contrast, LERGs are associated with a radiatively-inefficient accretion flow \citep[e.g.][]{1994ApJ...428L..13N,1995ApJ...452..710N,Yuan2014}, often modelled as arising from cooling hot gas, and as a result, do not display signs of a typical AGN such as an optically thick accretion disc or a torus, and therefore lack high-excitation emission lines in their optical spectra. In this scenario, the HERGs have been argued to typically accrete at $\gtrsim$ 1 per cent of the Eddington-scaled accretion rate, whereas the LERGs accrete at much lower rates of $\lesssim$ 1 per cent of the Eddington-scaled accretion rate \citep{2012MNRAS.421.1569B,Mingo2014}.

Studies of LERGs and HERGs in the nearby Universe have also shown differences in host galaxy properties between the two classes. LERGs are typically found to be hosted in massive, quiescent, red galaxies with massive black holes, and are often found in rich group or cluster environments, whereas HERGs tend to be found in less massive galaxies, with recent or on-going star-formation, and in poorer environments \citep[e.g.][]{Tasse2008,Smolcic2009,2012MNRAS.421.1569B,Gendre2013,Sabater2013,Mingo2014,Ching2017,Williams2015,Williams2018,Croston2019,Magliocchetti2022}. Although both LERGs and HERGs are found across a wide range of luminosities, the characteristic break in the luminosity function for HERGs occurs at higher luminosities, and they are found to also show a stronger space density evolution with redshift \citep[e.g.][]{2012MNRAS.421.1569B,2014MNRAS.445..955B,Pracy2016,Butler2019,Kondapally2022}. These observed properties of the host galaxies of LERGs and HERGs can be understood in terms of the different accretion properties of these AGN. The cooling of hot gas within massive, quiescent galaxies is expected to lead to low accretion rates resulting in the formation of a LERG, whereas the plentiful supply of cold gas present in lower mass, star-forming systems may also lead to higher accretion rates, resulting in a HERG. 

Using deep observations from the LOFAR Two-metre Sky Survey Deep Fields Data Release 1 (LoTSS-Deep DR1; \citealt{Tasse2021,Sabater2021,Kondapally2021,Duncan2021,Best2023}), \citet{Kondapally2022} studied the cosmic evolution of LERGs, finding that at $z \gtrsim 1$, most of the LERGs are hosted by star-forming galaxies, in contrast to studies at lower redshifts where LERGs are predominantly found in quiescent galaxies \citep[e.g.][]{2012MNRAS.421.1569B}. These results suggest that there may be differences in the host galaxy properties of AGN in the early Universe and at lower luminosities, with some overlap with the host galaxy properties of HERGs \citep[see also][]{Whittam2018,Whittam2022}. Moreover, the significant population of LERGs hosted in a different galaxy type (star-forming as opposed to quiescent) at early times can have interesting implications for both the fuelling mechanisms \citep[see also][]{Delvecchio2022} and our current understanding of AGN feedback processes and its effect on the galaxy population \citep[e.g.][]{cattaneo2009_feedback_review,smolcic2017agn_evol_vla,Butler2019,Hardcastle2020,Kondapally2023,Heckman2024}.

It is well known that in the local Universe, the fraction of galaxies hosting a radio-AGN (which provides a measure of the duty cycle) strongly increases with stellar mass as $f_{\rm{radio-AGN}} \propto M_{\star}^{2.5}$ \citep{Best2005fagn,Smolcic2009}, with recent work by \citet{sabater2019lotssagn} finding that at low luminosities almost all massive galaxies host a radio-AGN. \citet{Janssen2012} studied the dependence of radio-AGN activity on galaxy properties at $z < 0.3$ using the radio-AGN sample of \citet{2012MNRAS.421.1569B}. Using this sample, \citet{Janssen2012} measured the incidence of LERGs and HERGs, separately, as a function of stellar mass at $z < 0.3$, finding that LERG activity increases steeply with stellar mass, consistent with $M_{\star}^{2.5}$, whereas the HERGs showed a much shallower dependence of $\propto M_{\star}^{1.5}$. Moreover, they found that compared to red galaxies, HERG activity was significantly more enhanced in blue galaxies at fixed stellar mass. These results are consistent with the notion of LERGs undergoing accretion from cooling of hot gas within massive haloes, with the HERGs being fuelled by the abundant cold gas present in their lower mass (bluer), star-forming host galaxies \citep[e.g.][]{Best2005fagn}.

Early and recent works that extended this analysis out to higher redshifts have found a similarly steep stellar mass dependence for the entire radio-AGN population \citep[e.g.][]{Tasse2008,Smolcic2009,Simpson2013,Williams2015,Williams2018,Wang2024,Igo2024}. Since the radio-AGN population consists of both the LERGs and HERGs, \citet{Kondapally2022} used data from one of the LoTSS Deep Fields to study how the incidence of LERGs, alone, depends on stellar mass within quiescent and star-forming galaxies, separately, to $z \sim 1.5$. They found that LERGs hosted by quiescent galaxies showed a redshift-invariant steep stellar mass dependence, highly consistent with observations in the local Universe of fuelling occurring from cooling hot gas \citep[see also][]{Williams2018}. On the other hand, the LERGs hosted by star-forming galaxies showed a much shallower dependence on stellar mass, suggesting that LERGs in more star-forming systems may be fuelled in a manner different from those in quiescent galaxies.

In this paper, we aim to extend these analyses to understand how radio-AGN are triggered across the galaxy population by measuring the incidence of these radio-AGN as a function of each of stellar mass, star-formation rate, and redshift; using the full LoTSS Deep Fields dataset allows us to split our sample across these parameters simultaneously. This paper is structured as follows. Section~\ref{sec:data} describes the radio and multi-wavelength dataset, along with the selection of radio-AGN and the parent sample used for comparison. Section~\ref{sec:flerg_fherg_mass} presents the results on the incidence of LERGs and HERGs as a function of mass and star-formation rate. In Section~\ref{sec:lerg_sfr_ms_off} we present the results on the dependence of LERG activity as a function of star-formation rate relative to the main sequence of star-formation. In Section~\ref{sec:edd_dist}, we present the Eddington-scaled accretion rate properties of the LERGs and HERGs. In Section~\ref{sec:fuelling_discussion} we present the interpretation and discussion of our results. Section~\ref{sec:conclusions} presents the conclusions of our study. Throughout this work, we use a flat $\Lambda$CDM cosmology with $\Omega_{\rm{M}}=0.3,~\Omega_{\rm{\Lambda}}=0.7$ and $H_{0} = 70~\mathrm{km~s^{-1}~Mpc^{-1}}$, and a radio spectral index $\alpha = -0.7$ (where $S_{\nu} \propto \nu^{\alpha}$).

\section{Data}\label{sec:data}
\subsection{Radio and other multi-wavelength data}\label{sec:radio_opt_data}
The radio dataset used in this analysis comes from LoTSS-Deep DR1 \citep{Tasse2021,Sabater2021}, based on LOFAR High Band Antenna (HBA) observations ranging from $\sim$ 113 to 177\,MHz\footnote{We refer to the central frequency throughout this paper as 150\,MHz, for simplicity.}. The LoTSS-Deep DR1 consists of repeated observations of the ELAIS-N1, Lockman Hole, and Bo\"{o}tes fields, totalling 168, 112, and 80\,hours, and reaching an rms sensitivity of 20, 22, and 32\,$\rm{\mu Jy\,beam^{-1}}$ near the centre of each field, respectively. Radio source catalogues were extracted out to the 30 per cent power point of the primary beam in each field using Python Blob Detector and Source Finder (\textsc{PyBDSF}; \citealt{2015ascl.soft02007M}) as detailed by \citet{Tasse2021} and \citet{Sabater2021}.

The three fields benefit from extensive deep, wide-area multi-wavelength imaging, including photometry from the ultraviolet, optical, near and mid-infrared, and far-infrared wavelengths provided by a suite of ground and space-based imaging surveys over the past two decades; the full details of the available multi-wavelength imaging dataset are described by \citet{Kondapally2021}. Using this multi-wavelength dataset, multi-band forced, matched-aperture photometry catalogues were generated in ELAIS-N1 and Lockman Hole \citep{Kondapally2021}. In the Bo\"{o}tes field, an adapted version of the point spread function matched catalogues from \citet{2007ApJ...654..858B,2008ApJ...682..937B} were used. Photometric redshifts for the multi-band catalogues were determined by \citet{Duncan2021} using a hybrid method which combined standard template fitting methods with machine learning methods \citep[see][]{duncan2018photz_templates,duncan2018photz_ml}. For a small fraction of sources in each field, spectroscopic redshifts were available ($\sim$ 5, 5, and 20 per cent in ELAIS-N1, Lockman Hole, and Bo\"{o}tes, respectively) and were used instead.

The host-galaxy identification process was then carried out for the LOFAR detected sources using these multi-band catalogues as detailed by \citet{Kondapally2021}. In summary, host-galaxy counterparts were identified using the Likelihood Ratio method \citep{1977A&AS...28..211D,1992MNRAS.259..413S} for suitable sources (typically compact sources with well-defined positions), or otherwise using a visual classification scheme where identifications and radio-source associations were performed using consensus decisions from members of the LOFAR collaboration \citep[see also][]{2019A&A...622A...2W}. This source-association and cross-matching process resulted in a catalogue of 81\,951 radio sources with host-galaxies identified for $>$97 per cent of these. Overall, the LR method was used to identify counterparts for $\sim$ 83 per cent of the sources, with a reliability and completeness of over 99 per cent, with the remaining 14 per cent of sources being classified visually (see \citealt{Kondapally2021} for further details). In this paper, we restrict our analysis to sources within the redshift range $0.3 < z \leq 1.5$. We exclude sources with $z < 0.3$ as the aperture-corrected photometry, and hence the derived photometric redshifts, may not be as robust due to the extended nature of sources, while sources above $z = 1.5$ are excluded as the photometric redshifts for galaxy-dominated sources become less reliable beyond this redshift \citep[see][]{Duncan2021}.

\subsection{Identification of AGN}\label{sec:sclass}
Spectral energy distribution (SED) fitting can be used not only to derive galaxy properties (such as stellar masses, star-formation rates, etc.) but also to identify and characterise AGN activity \citep[e.g.][]{CalistroRivera2016,Leja2017,Boquien2019,Das2024}. \citet{Best2023} performed SED fitting for the radio-detected sources in LoTSS-Deep DR1 to broadly classify them into star-forming galaxies (SFGs) and the different types of AGN. Deep radio continuum surveys are expected to detect a variety of source populations; a combination of different SED fitting codes that can model AGN emission and those that allow a better sampling of normal galaxies (i.e. those without an AGN component) were used by \citet{Best2023} to optimise the classification process. In summary, for each radio source, using the best-estimate redshift (photometric or spectroscopic if available), SED fitting was performed using each of \textsc{agnfitter} \citep{CalistroRivera2016}, \textsc{bagpipes} \citep{Carnall2018}, \textsc{cigale} \citep{Burgarella2005,Noll2009,Boquien2019}, and \textsc{mahphys} \citep{daCunha2008}. Both \textsc{agnfitter} and \textsc{cigale}, unlike the other two codes, employ models for fitting the emission from the AGN accretion disc and torus which can result in more robust fits of physical properties for galaxies hosting an AGN, while also allowing the identification of AGN.

\citet{Best2023} began by identifying radiative-mode AGN using the outputs from the four SED fitting codes. Firstly, using the results from \textsc{agnfitter} and \textsc{cigale}, they defined a parameter $f_{\rm{AGN, 16}}$ (from each code) which represents the 16th percentile of the fraction of the total mid-infrared luminosity arising from AGN components. Then, the reduced $\chi^{2}$ values from \textsc{agnfitter} and \textsc{cigale}, which model AGN emission, were compared to the reduced $\chi^{2}$ values from \textsc{bagpipes} and \textsc{magphys}; for sources with infrared AGN emission, the former two codes should find a better SED fit. The combination of these two criteria were primarily used to identify the radiative-mode AGN (see \citealt{Best2023} for the exact criteria employed). In addition to the above, for ELAIS-N1 and Lockman Hole, bright X-ray detected AGN were identified using the Second \textit{ROSAT} All-Sky Survey (2RXS; \citealt{Boller2016}) and the \textit{XMM-Newton} Slew Survey (XMMSL2)\footnote{\url{https://www.cosmos.esa.int/web/xmm-newton/xmmsl2-ug}}. In Bo\"{o}tes, deep, wide-area observations from the X-Bo\"{o}tes survey \citep{Kenter2005} were used to identify AGN \citep[see][]{Duncan2021}. In addition, for a small fraction of the sources, optical spectroscopy indicated the presence of an AGN, which were also classified as radiative-mode AGN. The final list of radiative-mode AGN were comprised of those classified from the \citeauthor{Best2023} SED fitting, plus a small fraction of additional X-ray or spectroscopically selected AGN that were not already identified from the SED fits. For sources classified as radiative-mode AGN, the `consensus' SFRs and stellar masses were primarily derived using the \textsc{cigale} output (as \textsc{bagpipes} and \textsc{magphys} may be unreliable as they did not model the AGN component and \textsc{agnfitter} was found to result in larger uncertainties in general). For sources not classified as radiative-mode AGN, consensus SFRs and stellar masses were estimated using the mean of the values from \textsc{bagpipes} and \textsc{magphys} (provided an acceptable fit was found in each case) as these are expected to be most reliable in the absence of AGN activity due to their sampling of galaxy parameters. Throughout the analysis in this paper, we use these consensus SFRs and stellar masses for our radio-AGN sample.

A tight-correlation is observed between radio-luminosity and star-formation rate for star-forming galaxies \citep[e.g.][]{CalistroRivera2016,gurkan2018lofar_sfr,Smith2021,Das2024}, which can be used to identify radio-loud AGN as sources that show a significant radio-luminosity excess (also known as `radio-excess' AGN) relative to this relation \citep[e.g.][]{smolcic2017_vla_3ghz_counterparts,Delvecchio2017,Williams2018,Whittam2022}. Following this, \citet{Best2023} used a `ridgeline' approach, which is well-fitted by $\log_{10}(\rm{L_{150\,MHz}/W\,Hz^{-1}}) = 22.24 + 1.08 \log_{10}(\rm{SFR/M_{\odot}\,yr^{-1}})$, to select radio-excess AGN as sources that showed a radio-excess of $>0.7\,\rm{dex}$ ($\approx 3\sigma$) compared to that expected from star-formation alone.

In this study, HERGs are identified as the radiative-mode AGN that also display a radio-excess AGN. LERGs do not display signs of typical AGN accretion disc or torus features and are identified by the presence of radio-jets only; these sources are therefore selected as sources that show a radio-excess AGN but are not classified as radiative-mode AGN. In total, across the redshift range $0.3 < z \leq 1.5$, there are 38,915 radio-detected sources, of which 7530 are radio-excess AGN, which are further split into 6775 LERGs and 755 HERGs.

\subsection{Potential incompleteness of radio-excess AGN at high star-formation rates}\label{sec:re_incomp}
The use of a radio-excess criterion in selecting radio-AGN may miss low-luminosity AGN, which may potentially bias our subsequent analysis. In particular, for sources with a given jet (AGN) luminosity, it will be more difficult to identify AGN using their radio-excess in more highly star-forming galaxies, leading to some incompleteness of radio-excess AGN at high star-formation rates. This may introduce a bias to our results, in conjunction with the application of a radio luminosity limit ($L_{\rm{150\,MHz}} \geq 10^{24}\,\rm{W\,Hz^{-1}}$; see Sect.~\ref{sec:flerg_fherg_mass}). The potential impact of this selection effect on our results could be mitigated by imposing a higher radio luminosity limit, however, this would result in poorer statistics and limit our sample to only the most luminous radio-AGN whose results may not be applicable to the broader radio-AGN population. In this paper, we instead perform Monte Carlo simulations to assess and correct for the impact of this selection effect on our results; these are described in detail in Appendix~\ref{ap:re_comp}. In summary, we simulate the total 150\,MHz radio luminosity of sources as arising from the sum of a star-formation component and an AGN (jet) component, while accounting for the radio-luminosity -- SFR relation and its scatter. The fraction of the simulated sources that would satisfy the radio-excess criterion of \citet{Best2023} was calculated to determine the completeness in narrow bins in radio luminosity and star-formation rate. For each star-formation rate bin, the completeness was weighted by the luminosity functions of various AGN types used in this study, down to the luminosity limit of $L_{\rm{150\,MHz}} \geq 10^{24}\,\rm{W\,Hz^{-1}}$ to compute the completeness corrections for the radio-excess selection. The output of this analysis is a set of completeness corrections as a function of SFR for different types of AGN (see Fig.~\ref{fig_ap:re_sfr_comp}). The results of this simulation show that our sample is consistent with 100 per cent completeness at $\rm{SFR} \lesssim 10\,\rm{M_{\odot}\,yr^{-1}}$, with this decreasing steeply at higher SFRs. The completeness corrections derived from these simulations are used for further analysis in Section~\ref{sec:flerg_fherg_mass} and~\ref{sec:lerg_sfr_ms_off}.

\subsection{Mid-infrared flux-selected parent sample}\label{sec:parent_sample}
To place the properties of the radio-AGN within the broader context of the galaxy population, one must compare to a sample of underlying/parent galaxies. The parent sample used for this comparison comes from the IRAC 3.6\,$\mu$m flux selected ($\rm{F_{\rm{3.6\,\mu m}}} > 10\,\rm{\mu Jy}$) sample taken from the multi-wavelength catalogues of \citet{Kondapally2021}; SED fitting for this sample was performed using \textsc{magphys} by \citet{Smith2021} to derive stellar masses and SFRs. \citet{Smith2021} presented these results only for the ELAIS-N1 field (the deepest of the LoTSS Deep Fields), however this process was subsequently repeated for both the Lockman Hole and Bo\"{o}tes fields (Smith et al. priv. communication) using similar input multi-wavelength photometry catalogues from \citet{Kondapally2021} which we use in this study. As noted by \citet{Duncan2021}, the photometric redshifts for galaxy-dominated (rather than AGN-dominated) systems in the three LoTSS Deep Fields are found to be most robust out to $z \sim 1.5$, and therefore the SED fitting process for the IRAC-selected sample (and therefore for the LoTSS-Deep sample), and further analysis in this paper is also limited to $z \leq 1.5$. In total, across the redshift range $0.3 < z \leq 1.5$, there are 540,279 sources in the parent IRAC-selected sample. Throughout this analysis, this parent MIR-selected sample (and the resulting physical properties such as SFRs and stellar masses) is used for comparison with the LoTSS-Deep radio-AGN sample. As shown by \citet{Best2023}, the stellar masses and SFRs derived for the parent sample, although using a slightly different method compared to the LoTSS Deep sources, are consistent.

Throughout the analysis in this paper, we compare the properties of the radio-AGN as a function of their stellar mass. For this analysis, it is important to account for mass incompleteness in our sample to avoid any biases in the interpretation of our results. We use the stellar mass completeness limits determined by \citet{Duncan2021} for the underlying multi-wavelength dataset which was used to identify the host galaxies of the radio sources and forms the basis of the mid-infrared parent sample. \citet{Duncan2021} determined the 90 per cent stellar mass completeness limits as a function of redshift by accounting for a maximally old stellar population, ensuring a complete sample of quiescent galaxies for each of the three LoTSS Deep fields, separately. Using these completeness limits, in the analysis throughout the rest of this paper, we estimate the mass above which a source would be detected over an entire chosen redshift range and remove sources (from both the radio-AGN and the mid-infrared parent sample) below this mass limit from the analysis.

\begin{figure*}
    \centering
    \begin{subfigure}{0.5\textwidth}
    \centering
    \includegraphics[width=\textwidth]{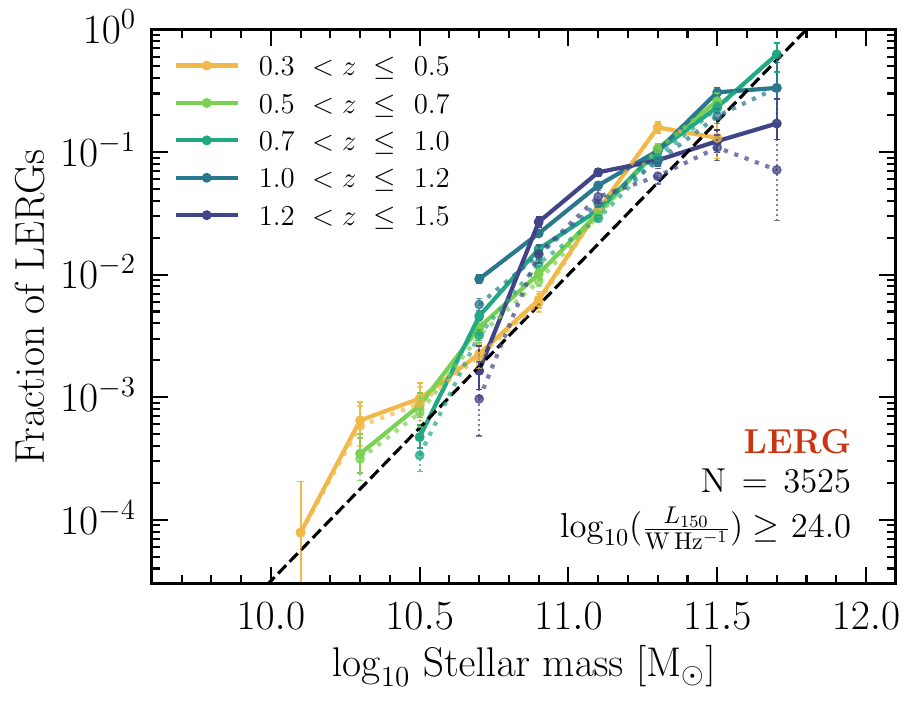}
    \end{subfigure}%
    \begin{subfigure}{0.5\textwidth}
    \includegraphics[width=0.983\textwidth]{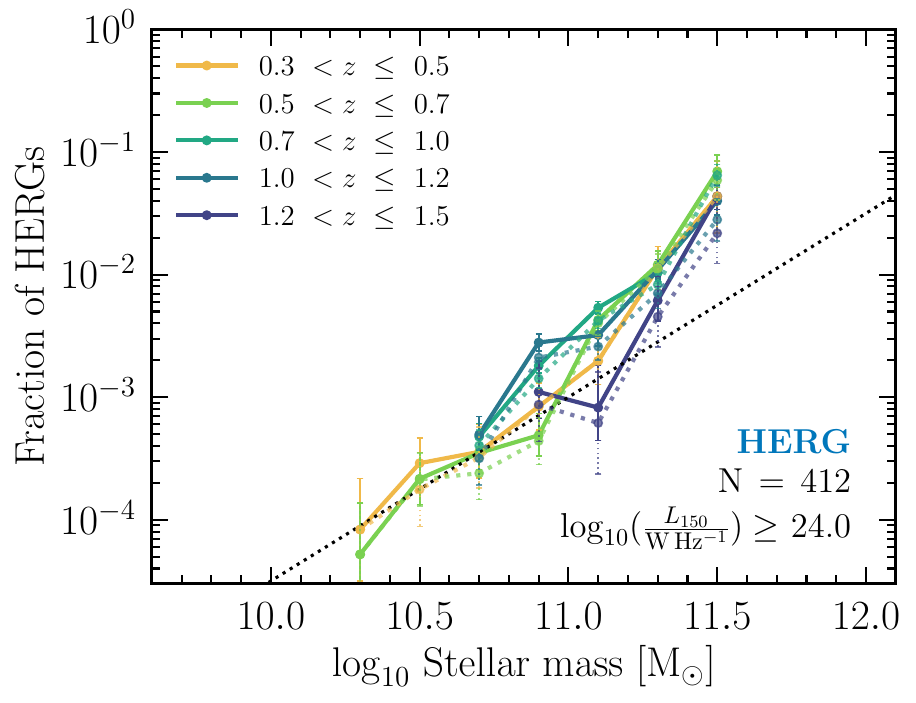}
    \end{subfigure}
    \caption{\label{fig:lerg_herg_frac}The fraction of all galaxies that host a LERG (\textit{left}) or a HERG (\textit{right}) as a function of stellar mass across $0.3 < z \leq 1.5$ for a radio luminosity limit of $L_{\rm{150\,MHz}} \geq 10^{24}\,\rm{W\,Hz^{-1}}$. The numbers of LERGs and HERGs satisfying the radio luminosity limit over this redshift range are also listed in each panel. The solid and dotted lines show the fractions obtained with and without the radio-luminosity -- SFR completeness corrections, respectively (see text and Appendix~\ref{ap:re_comp}); these corrections have a small effect for the overall populations. The black dashed line in the left panel represents the steep stellar mass relation ($\propto M_{\star}^{2.5}$, normalised to 0.01 at $M = 10^{11}\,\rm{M_{\odot}}$) found for LERGs in the local Universe \citep{Best2005fagn,Janssen2012}. The black dotted line in the right panel corresponds to the shallow mass dependence ($\propto M_{\star}^{1.5}$, normalised to 0.001 at $M = 10^{11}\,\rm{M_{\odot}}$), similar to that found for HERGs in the local Universe \citep{Janssen2012}. The LERGs show the same steep relation as observed in the local Universe all the way out to $z \sim 1.5$. The HERG fraction at low masses follow a shallower relation at $M < 10^{11}\,\rm{M_{\odot}}$, but show a steeper than expected relation (based on higher frequency local Universe studies) at higher masses.}
\end{figure*}

\section{The incidence of radio-AGN activity on stellar mass and SFR}\label{sec:flerg_fherg_mass}
In Fig.~\ref{fig:lerg_herg_frac}, we present the cosmic evolution of the fraction of all galaxies that host a LERG (\textit{left}) or a HERG (\textit{right}) as a function of stellar mass for a radio luminosity limit of $L_{\rm{150\,MHz}} \geq 10^{24}\,\rm{W\,Hz^{-1}}$. The results are calculated across five redshift bins ($0.3 < z \leq 0.5$, $0.5 < z \leq 0.7$, $0.7 < z \leq 1.0$, $1.0 < z \leq 1.2$, $1.2 < z \leq 1.5$), shown by the different colours. The error bars are calculated following binomial statistics. The chosen radio luminosity limit of $L_{\rm{150\,MHz}} \geq 10^{24}\,\rm{W\,Hz^{-1}}$ corresponds to a 5$\sigma$ detection limit (based on the depth of the deepest field) for a source out to $z \sim 1.5$; moreover, this 150\,MHz radio luminosity limit is also comparable to the 1.4\,GHz radio luminosity limit of $10^{23}\,\rm{W\,Hz^{-1}}$ often used in the literature for similar studies \citep[e.g.][]{Best2005fagn,Janssen2012}, given the typical radio spectral index (assuming $\alpha \approx -0.7$). There are 3525 LERGs and 412 HERGs that satisfy the above radio luminosity limit across the chosen redshift range for our analysis. In the left panel for the LERGs, the black dashed line represents the fraction of galaxies hosting a LERG at a given stellar mass, $f_{\rm{LERG}}$, found in the local Universe, given as $f_{\rm{LERG}} = 0.01 \left(M_{\star}/10^{11}\,M_{\odot}\right)^{2.5}$ \citep{Best2005fagn,Janssen2012}; this relation has been roughly normalised based on these low redshift results but we note that the precise normalisation depends on the radio luminosity limit and hence on the assumed spectral index. In the right panel, the black dotted line corresponds to the shallower stellar mass dependence found for HERGs in the local Universe ($\propto M_{\star}^{1.5}$; \citealt{Janssen2012}), which has been scaled to correspond to $f_{\rm{HERG}} = 0.001 \left(M_{\star}/10^{11}\,M_{\odot}\right)^{1.5}$.

As noted in Sec.~\ref{sec:re_incomp}, we have derived completeness corrections to account for radio-excess selection effects, which are detailed in Appendix~\ref{ap:re_comp}; this analysis indicates that we achieve a lower completeness of radio-excess AGN at higher star-formation rates. For each stellar mass bin, we apply the corrections (for the LERGs and HERGs, separately) shown in Fig.~\ref{fig_ap:re_sfr_comp}. For each redshift bin in Fig.~\ref{fig:lerg_herg_frac}, the solid and dotted lines show the fractions with and without these completeness corrections applied, respectively. Applying this correction increases the fraction of galaxies hosting a LERG or a HERG, which is noticeable as we tend toward high stellar masses and high redshifts, where the average SFRs increase. For the LERGs, these corrections only significantly affect the most massive systems within the highest redshift bin in Fig.~\ref{fig:lerg_herg_frac}, however the data points with and without the corrections are consistent within their uncertainties. For the HERGs, small differences are seen across a wider range of stellar masses and redshifts. This suggests that the completeness corrections have a small effect on the stellar mass dependencies, particularly at high masses and redshifts.

\begin{figure*}
    \centering
    \includegraphics[width=0.9\textwidth]{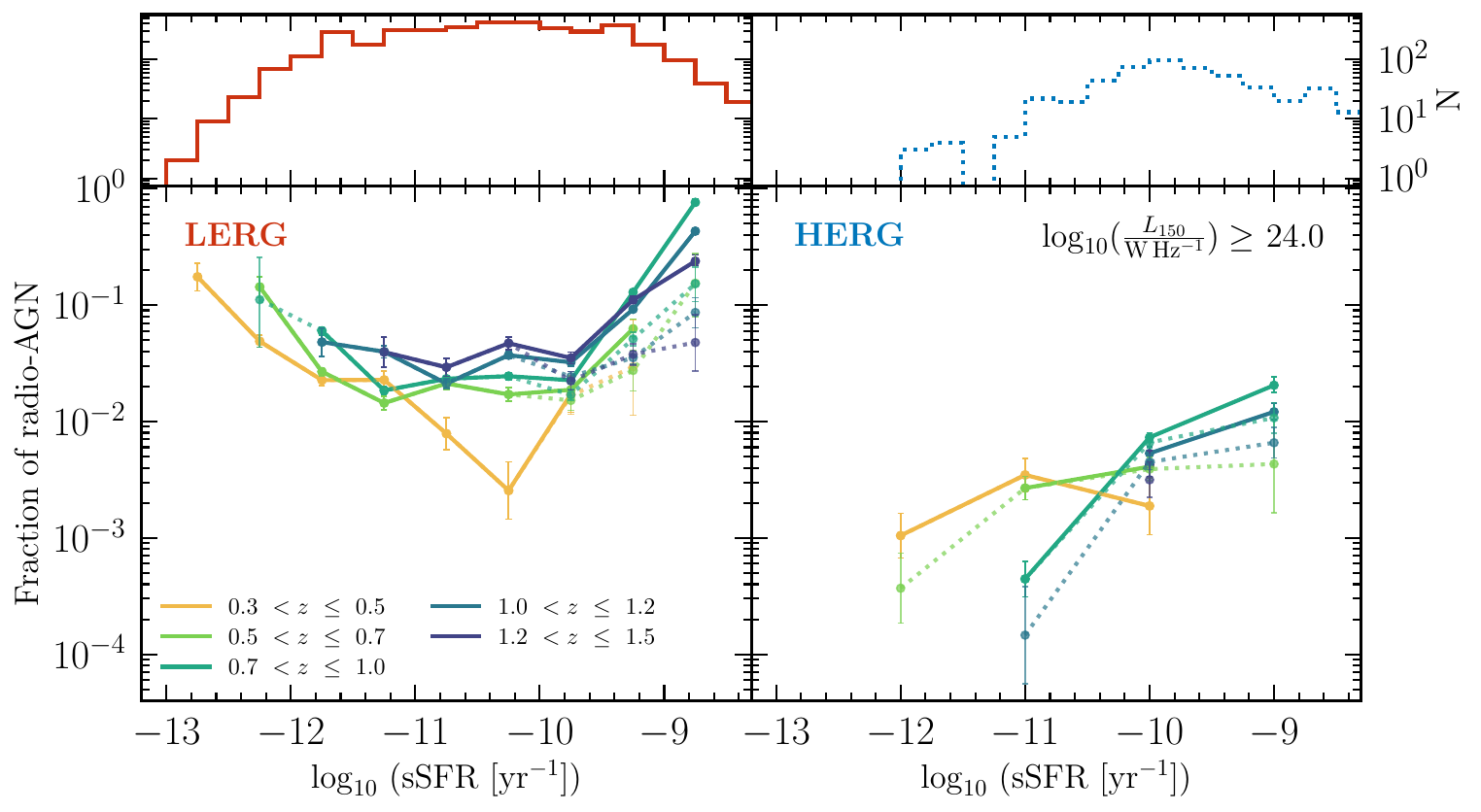}
    \caption{\label{fig:lerg_herg_ssfr_frac}The fraction of galaxies that host a LERG (\textit{lower left}) or a HERG (\textit{lower right}) with $L_{\rm{150\,MHz}} \geq 10^{24}\,\rm{W\,Hz^{-1}}$ as a function of sSFR across $0.3 < z \leq 1.5$. To reduce any mass selection effects, we calculate the fractions for a constant stellar mass range of $10.8 < \log_{10} (M_{\star}/M_{\odot}) \leq 11.5$ (see text) across the redshift bins. The solid and dotted lines show the results with and without the radio luminosity -- SFR completeness correction (see Sect.~\ref{sec:re_incomp}). The corresponding top panels on the left and right show the distribution of the sSFRs for the LERGs and HERGs, respectively. The LERGs show a broad sSFR distribution, whereas the HERGs tend to be preferentially hosted in star-forming galaxies. For the LERGs, the radio-AGN fraction remains mostly flat across many orders of magnitude in sSFR, only increasing above $\log_{10}(\rm{sSFR}/\rm{yr^{-1}}) \gtrsim -10$.}
\end{figure*}

The results from Fig.~\ref{fig:lerg_herg_frac} for the LERGs show that the incidence of LERGs has a steep stellar mass dependence out to $z \sim 1.5$. The dependence on stellar mass appears to slightly flatten with increasing redshift, as previously shown by \citet{Kondapally2022} for the LERGs and by \citet{Williams2015} for the total radio-AGN population. We note that the requirement of high black hole masses $(\rm{M_{BH}} > 10^{7.8}\,M_{\odot})$ to launch powerful radio jets found by \citet{Whittam2022} is consistent with the high prevalence of LERG activity in the most massive galaxies observed in this study. Broadly the results for the LERG population are consistent with studies in the local Universe \citep[e.g.][]{Best2005fagn,Tasse2008,Janssen2012,sabater2019lotssagn}, with some evidence of flattening at lower masses and higher redshifts, which we investigate further in Sect.~\ref{sec:lerg_sfr_ms_off}. The dependence of HERG activity on stellar mass shows different trends compared to local Universe studies. At $\log_{10} (M_{\star}/M_{\odot}) \lesssim 11$, the HERGs trace the slope of the black dotted line well, which corresponds to the shallow $\propto M_{\star}^{1.5}$ dependence observed in the local Universe, albeit with a higher normalisation by a factor of $\sim 2 - 3$ at a given stellar mass. \citet{Janssen2012} also studied the HERG fractions when split by those hosted by blue galaxies alone; their normalisation of the HERG fractions within blue galaxies agrees well with our results for the total HERG population shown here. At higher masses, and also with increasing redshift, our results indicate a steeper mass dependence, with a higher prevalence of HERGs in massive galaxies than expectations from low redshift studies.

The HERG population is known to exhibit strong evolution with redshift \citep[e.g.][]{2014MNRAS.445..955B,Pracy2016,Williams2018,Kondapally2022}; some of the observed differences with our study could be due to cosmic evolution, since even the lowest redshift bin in this study probes earlier epochs than the typical redshift of the \citet{Janssen2012} sample of $z_{\rm{med}} \sim 0.15$. Another source of differences could arise from the different source classification methods used to identify LERGs and HERGs. \citet{Janssen2012} used spectroscopic classifications by \citet{2012MNRAS.421.1569B} based on emission line ratio diagnostics to identify LERGs and HERGs; this is typically considered to be the `gold standard' method of classifying radiatively efficient versus inefficient accretion. In this paper, we do not have spectroscopic information available for the vast majority of the radio sources, therefore SED fitting is used for the classification instead (see Sect.~\ref{sec:sclass}). It is possible that a small fraction of LERGs mis-classified as HERGs could cause an increase in the AGN fraction of HERGs, particularly at high masses where the LERG fraction is $\sim$ 10 per cent. We discuss the potential uncertainties in our source classification method in Sect.~\ref{sec:sclass_edd} and further validate the robustness of this in Appendix~\ref{ap:ap_sclass}.

\begin{figure*}
    \centering
    \includegraphics[width=0.9\textwidth]{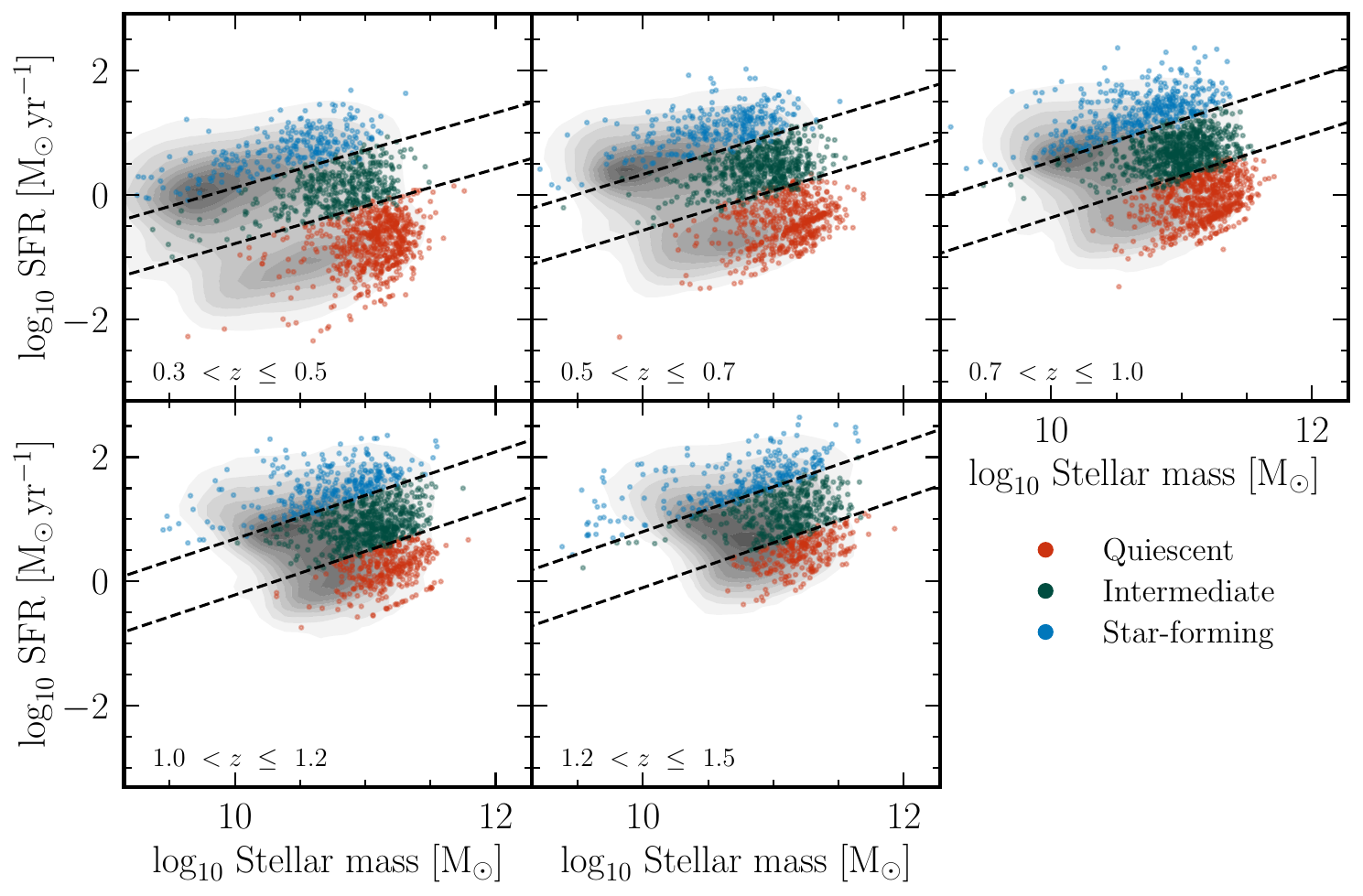}
    \caption{\label{fig:sfr_ms_lerg}The location of LERGs on the SFR-M$_{\star}$ plane in five redshift bins across $0.3 < z \leq 1.5$. The LERG sample has been split into the subset that are hosted by quiescent (red), intermediate (green), and star-forming (blue) galaxies, defined based on the evolving main-sequence (see Sec.~\ref{sec:lerg_sfr_ms_off}). The diagonal dashed lines show the division lines used to select the three populations. The shaded contours show the distribution of the underlying MIR flux-selected parent sample along the SFR-M$_{\star}$ plane.}
\end{figure*}

We investigate the dependence of radio-AGN activity on star-formation activity in Fig.~\ref{fig:lerg_herg_ssfr_frac}, which shows the fraction of galaxies that host a LERG (\textit{lower left}) or a HERG (\textit{lower right}) as a function of their specific star-formation rate ($\rm{sSFR = SFR / M_{\star}}$) for AGN with $L_{\rm{150\,MHz}} \geq 10^{24}\,\rm{W\,Hz^{-1}}$. The analysis is performed out to $z \sim 1.5$ over the same redshift bins as in Fig.~\ref{fig:lerg_herg_frac}. To ensure the results are not biased by selection effects across different redshift bins, we restrict this analysis to sources with $10.8 < \log_{10} (M_{\star}/M_{\odot}) \leq 11.5$; the lower limit is chosen to roughly corresponds to the 90 per cent stellar mass completeness limit at $z=1.5$. As in Fig.~\ref{fig:lerg_herg_frac}, the solid and dotted lines show the results with and without the radio-excess completeness corrections applied, respectively (see Appendix~\ref{ap:re_comp}). The uncertainties are calculated following binomial statistics. The top panels show the corresponding distributions of the sSFRs for the LERGs and HERGs.

The fraction of galaxies hosting a LERGs reaches $\gtrsim$ 10 per cent in the most quiescent galaxies (i.e. at the lowest sSFRs), before declining to a few per cent level across $-12 < \log_{10}(\rm{sSFR}/\rm{yr^{-1}}) < -10$. At higher sSFRs, for the most star-forming systems, the LERG fraction shows a strong increase; this trend is seen across redshift out to $z \sim 1.5$. We note the the dip in the fraction at intermediate sSFRs at $z \leq 0.5$ is only at a $\approx 2\sigma$ level with a small number of sources. Our results for the LERGs show that their hosts span a broad range of star-formation activities and that the prevalence remains broadly constant across a wide range of sSFRs; we investigate the trends with star-formation activity in more detail in Sect.~\ref{sec:lerg_sfr_ms_off}. The HERGs show a much lower prevalence than the LERGs at all sSFRs, similar to the stellar mass dependence results shown in Fig.~\ref{fig:lerg_herg_frac}. The results in Fig.~\ref{fig:lerg_herg_ssfr_frac} show that HERGs are more likely to be found in star-forming systems compared to quiescent systems, albeit with relatively small numbers of HERGs in each bin resulting in large statistical uncertainties. These findings for the HERGs are in good agreement with previous studies \citep[e.g.][]{2012MNRAS.421.1569B,Mingo2014,2014MNRAS.445..955B,Pracy2016}.

\section{Dependence of radio-AGN activity on SFR relative to the main sequence}\label{sec:lerg_sfr_ms_off}
In this section, we explore the properties of the LERGs in more detail by studying the incidence of LERGs as a function of their position relative to the star-forming main sequence. As the HERGs are typically found to be hosted in star-forming systems and given the relatively small numbers of HERGs observed, performing detailed statistical analyses for the HERGs as a function of SFR and stellar mass, simultaneously, is not feasible; we therefore focus our further analysis in this section on the LERGs alone.

We divide our AGN and galaxy samples into three sub-groups based on their SFR relative to the star-forming main sequence: quiescent, intermediate, and star-forming. For this, we use the best-fit relation for the star-forming main sequence from \citet{speagle2014sfrms}, which was found to provide a good fit to the observed SFRs out to $z \sim 5$, given as
\begin{equation}\label{eq:msrel}
\log_{10}(\mathrm{SFR_{MS}}(t))  = (0.84 - 0.026 \times t) \log_{10}(M_{\star}) - 6.51 + 0.11 \times t,
\end{equation}
where $\rm{SFR_{MS}}$ is the main-sequence star-formation rate (in $\rm{M_{\odot}\,yr^{-1}}$ units) and $t$ is the age of the Universe at $z$. Using this, we split our sample into three regions along the $\rm{SFR} - M_{\star}$ plane as follows:
\begin{enumerate}
    \item `Star-forming galaxies', as in galaxies that are on or above the main-sequence with $\log \left( \rm{SFR} / \rm{SFR_{MS}}(t) \right) \geq -0.4$.
    \item `Intermediate galaxies', as in galaxies that are below the main-sequence but with SFRs above that of a quiescent galaxy with $-1.3 < \log \left( \rm{SFR} / \rm{SFR_{MS}}(t) \right) \leq -0.4$.
    \item `Quiescent galaxies' with $\log \left( \rm{SFR} / \rm{SFR_{MS}}(t) \right) < -1.3$.
\end{enumerate}

The typical scatter in the star-forming main sequence relation derived by \citet{speagle2014sfrms} is $\sim$ 0.2\,dex. Following this, our selection of star-forming galaxies corresponds to sources within $\approx 2\sigma$ or above the main-sequence, with quiescent galaxies defined as sources that are at least $\approx 6.5\sigma$ below the main-sequence, and `intermediate' galaxies lying in between these two populations. By defining our sample based on SFRs relative to the main sequence across different redshift bins, we account for the evolution of the galaxy population when comparing results across redshift bins as galaxies typically have higher SFRs at higher redshifts. Moreover, this analysis is also less affected by any systematic uncertainties on SFR estimates from photometry, although this is expected to be small \citep[see][]{Best2023}.

In Fig.~\ref{fig:sfr_ms_lerg}, we show the SFR - $M_{\star}$ distribution of the LERGs hosted by star-forming galaxies (blue), intermediate galaxies (green), and quiescent galaxies (red) across $0.3 < z \leq 1.5$ over five redshift bins, as in Fig.~\ref{fig:lerg_herg_frac}. The two diagonal dashed lines in each redshift bin denote the criteria used to separate the sources into the above three galaxy types relative to the main sequence. The grey shaded region in each redshift bin corresponds to the distribution of the mid-infrared flux-selected parent sample in the SFR - $M_{\star}$ plane (as described in Sect.~\ref{sec:parent_sample}) which was used for calculating the incidence of AGN in this section.

\begin{figure*}
    \centering
    \includegraphics[width=\textwidth]{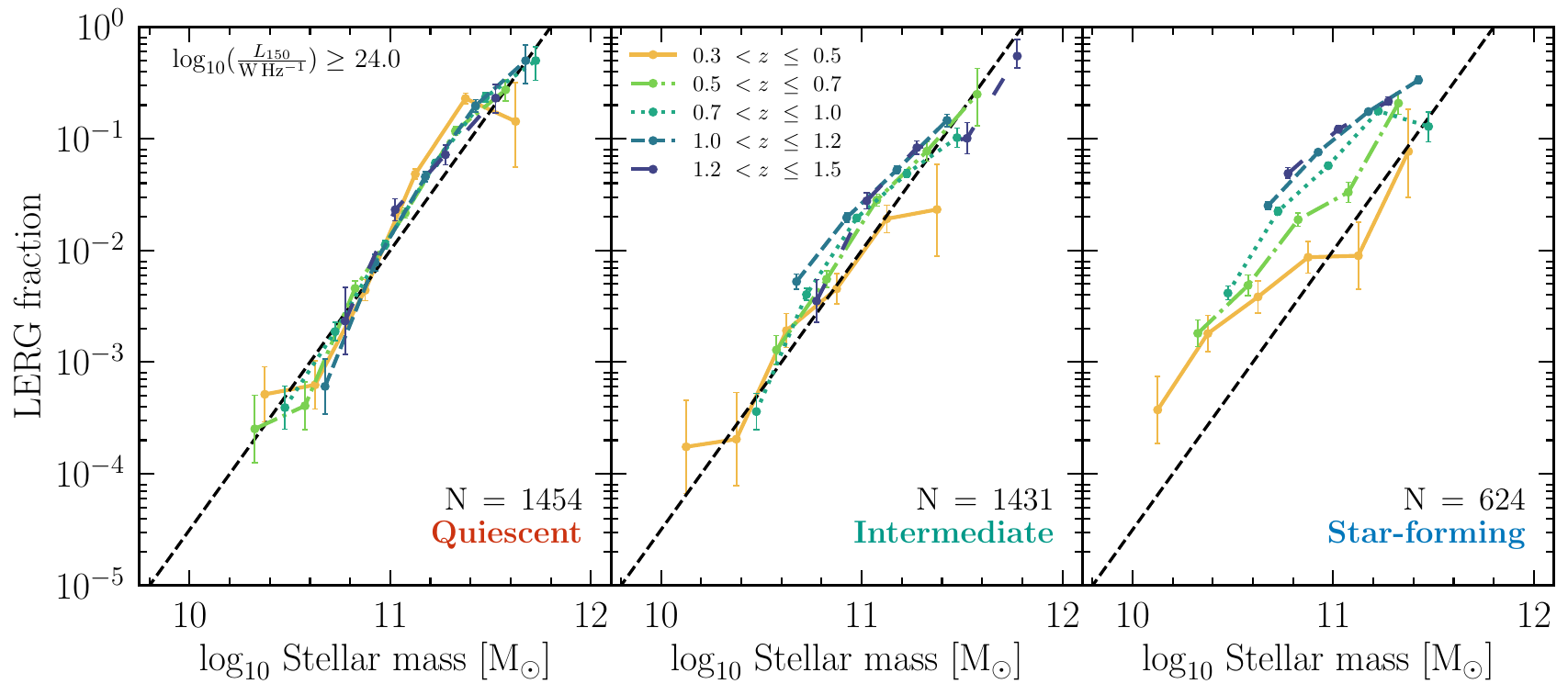}
    \caption{\label{fig:lerg_frac_sfr_ms}Fraction of quiescent (left), intermediate (middle), and star-forming (right) galaxies hosting a LERG with $L_{\rm{150\,MHz}} \geq 10^{24}\,\rm{W\,Hz^{-1}}$ as a function of their stellar mass across $0.3 < z \leq 1.5$. The host galaxy types for the LERGs are identified based on the offset from the evolving  star-forming main sequence, with the number in each panel corresponding the number of LERGs hosted by each galaxy type across $0.3 < z \leq 1.5$ for the above radio luminosity limit. The LERGs hosted by quiescent galaxies shows a steep dependence on stellar mass, that does not evolve with cosmic time; moving along the galaxy population, towards more star-forming systems, the incidence of LERGs increases at lower masses leading to a flattening of the relationship.}
\end{figure*}

In Fig.~\ref{fig:lerg_frac_sfr_ms} we compute the fraction of quiescent (left), intermediate (middle), and star-forming (right) galaxies that host a LERG with $L_{\rm{150\,MHz}} \geq 10^{24}\,\rm{W\,Hz^{-1}}$ as a function of stellar mass across $0.3 < z \leq 1.5$. The uncertainties are calculated following binomial statistics. In each panel, the black dashed line shows the local Universe relationship $f_{\rm{LERG}} = 0.01 \log_{10}(M_{\star}/10^{11}\,\rm{M_{\odot}})^{2.5}$ for the LERGs \citep{Janssen2012}. The results shown in this figure extend beyond the analysis performed by \citet{Kondapally2022} of the binary separation of star-forming and quiescent galaxies.

\begin{figure*}
    \centering
    \includegraphics[width=\textwidth]{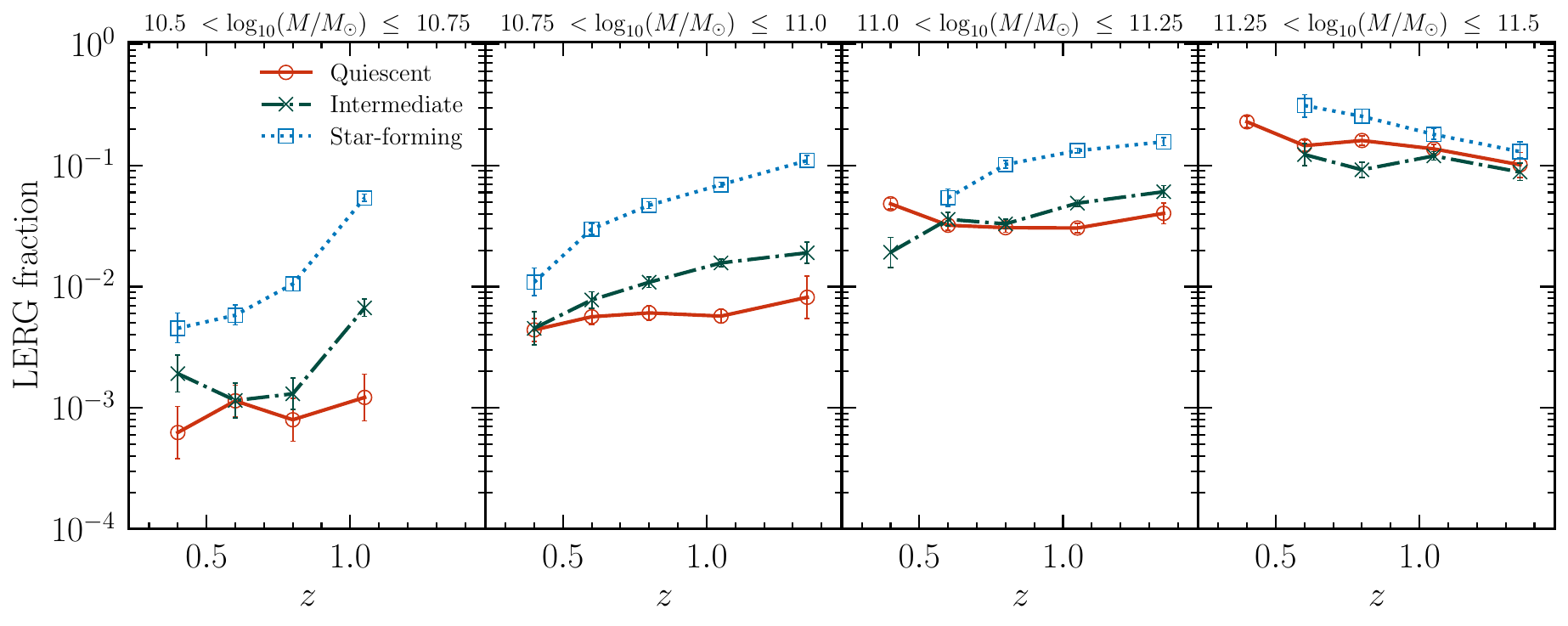}
    \caption{\label{fig:lerg_frac_sfr_ms_mass}The incidence of LERGs with $L_{\rm{150\,MHz}} \geq 10^{24}\,\rm{W\,Hz^{-1}}$ across redshift, with panels showing increasing stellar mass bins from left to right. In each panel, the incidences of LERGs are calculated within different galaxy types: quiescent (red), intermediate (green), and star-forming (blue). Overall the incidence of LERGs increases with stellar mass across all galaxies, going from the left to the right panel. At low masses of $\log_{10}(M/M_{\odot}) \leq 11$, galaxies with ongoing star-formation activity (i.e. the intermediate and star-forming populations) are significantly more likely to host a LERG compared to quiescent galaxies, across all redshifts studied. At higher masses, the LERGs are found across the entire galaxy population, regardless of their star-formation activity, at roughly the same rate.}
\end{figure*}

Across the five redshift bins (shown by different coloured lines, as in Fig.~\ref{fig:lerg_herg_frac}), the LERGs hosted by quiescent galaxies (Q-LERGs hereafter) follow the same steep dependence on stellar mass with the same normalisation out to $z \sim 1.5$; this is consistent with the observations of the overall LERG population in the local Universe, suggesting that LERGs within quiescent galaxies are also fuelled by the same mechanism, the cooling hot gas, right across cosmic time. As the star-formation activity increases, going from intermediate to the star-forming populations in Fig.~\ref{fig:lerg_frac_sfr_ms}, the dependence on stellar mass flattens due to an increase in the incidence of LERGs within lower mass systems; as a result, the star-forming LERGs (SF-LERGs) are no longer consistent with the steep relation observed for the quiescent hosts, in particular at low masses. Moreover, as the fractions are calculated over the given galaxy type (rather than the entire galaxy population), the increase in the incidence of LERGs in SFGs at higher redshifts for fixed stellar mass indicates that SFGs are more likely to host a LERG at higher redshifts than in the local Universe. One potential cause for this result may be due to the increase in gas fractions at higher redshifts \citep[e.g.][]{Genzel2015,Tacconi2018}; the more abundant gas supply that is available to fuel the more vigorous star-formation within these host galaxies may also fuel the AGN.

\citet{Janssen2012} studied the dependence of LERG activity on galaxy colour at $z < 0.3$ using D4000{\AA} and \textit{u - r} colours to separate sources into red, green, and blue galaxies. They found that compared to green or blue galaxies, red galaxies were  more likely to host a LERG at high masses (by a factor of $\sim$ 2- 3 times at $\log_{10}(M_{\star}/M_{\odot}) \gtrsim 11$), with the incidence across galaxies of different colours becoming more similar at low masses. However our results indicate that at high masses the incidence of LERGs across quiescent and star-forming galaxies becomes roughly similar. Some of the differences with \citet{Janssen2012} could be due to cosmic evolution; only $\sim$ 2 per cent of the LERGs identified by \citet{Janssen2012} were hosted by blue galaxies whereas LERGs are found to be more likely to be hosted by star-forming galaxies with increasing redshift \citep[see][]{Kondapally2022}. We also note that while their separation of galaxies into red, green, and blue is analogous, it is not directly comparable to the quiescent, intermediate, and star-forming separation applied in this study.

\citet{Williams2015} studied the incidence of radio-loud AGN on stellar mass out to $z \sim 2$ using early LOFAR observations. They found that the radio-loud AGN fraction (i.e. for LERGs and HERGs combined) increased at low masses with increasing redshift, while staying roughly constant at high masses; this is in good agreement with our results. Compared to their results for the entire radio-loud population, we have shown that the flattening of the mass dependence at high redshifts is likely caused by an increased prevalence of LERGs hosted by star-forming galaxies. Recently, \citet{Wang2024} studied the incidence of radio-excess AGN in quiescent and star-forming galaxies out to higher redshifts. They find that radio-excess AGN are more likely to be found in massive quiescent galaxies compared to star-forming galaxies, which is broadly consistent with our results. Their analysis is carried out for a sample of all radio-excess AGN, which includes both the LERG and HERG populations, and is performed in bins of radio luminosity rather than integrated over all radio luminosities above a limit (as in this study); this prevents a more quantitative comparison with our results.

To better illustrate the stellar mass and any redshift dependence across different galaxy populations, in Fig.~\ref{fig:lerg_frac_sfr_ms_mass}, we show the fraction of different galaxy types hosting a LERG (with $L_{\rm{150\,MHz}} \geq 10^{24}\,\rm{W\,Hz^{-1}}$) as a function of redshift within four stellar mass bins (separated by 0.25\,dex) increasing in mass from left to right panels. In the lowest stellar mass bins with $10.5 < \log_{10}(M/M_{\odot}) \leq 11$, there is a notable difference in the incidence of LERGs with increasing star-formation activity. The fraction of LERGs hosted by quiescent galaxies is typically lower by an order of magnitude or more compared to the star-forming galaxies. The fraction of LERGs also increases with increasing redshift (by a factor of $\sim$ 2 - 8 out to $z \sim 1$) for intermediate and star-forming galaxies, whereas this remains largely flat for the quiescent galaxies. At $10.75 < \log_{10}(M/M_{\odot}) \leq 11$, the star-forming galaxies are still more likely to host a LERG than quiescent galaxies at any redshift, with the LERG fraction increasing with increasing redshift for the more star-forming systems. The higher prevalence of LERG activity in star-forming galaxies compared to quiescent galaxies, coupled with the increase in the LERG fraction within star-forming galaxies with increasing redshift, results in the overall flattening of the relation seen for the star-forming LERGs in Fig.~\ref{fig:lerg_frac_sfr_ms}. These results at low masses suggest that there is a causal link between the triggering of LERG activity and the cold gas fractions in star-forming galaxies, which increases at higher redshifts \citep[e.g.][]{Genzel2015,Tacconi2018}. In the most massive bin, the incidence of LERGs remains largely consistent across different galaxy types: at $11.25 < \log_{10}(M/M_{\odot}) \leq 11.5$, $\sim$ 10 -- 20 per cent of all galaxies host a LERG, regardless of their star-formation activity. This suggests that LERG activity in massive galaxies may be driven by the same mechanism across the galaxy population: cooling of hot gas from their haloes. Although the fraction of star-forming and quiescent galaxies hosting a LERG are similar at these high masses, there are more star-forming galaxies at higher redshifts than quiescent galaxies, so the bulk of the AGN activity in the early Universe occurs in galaxies that are not quiescent \citep[see also][]{Kondapally2022}.

\section{Distribution of the Eddington-scaled accretion rates}\label{sec:edd_dist}
Observations of different states of X-ray binaries 
\citep[e.g.][]{Remillard2006} and theoretical accretion disc models \citep[e.g.][]{1994ApJ...428L..13N,1995ApJ...452..710N,Yuan2014} indicate a switch in the nature of the accretion flow occurring at Eddington-scaled accretion rates, $\lambda_{\rm{Edd}}$, of around 1 per cent ($\lambda_{\rm{Edd}} = 0.01$), going from radiatively efficient accretion at higher accretion rates to radiatively inefficient (or advection dominated accretion) at lower accretion rates. In the local Universe, studies have previously found significant differences in the Eddington-scaled accretion rates of the LERGs and HERGs, with the LERGs typically accreting below 1 per cent of the Eddington rate, and the HERGs typically accreting at higher rates \citep[e.g.][]{2012MNRAS.421.1569B,Mingo2014}. These differences in the accretion rate properties have been used to explain many of the differences in the properties of the central AGN and their host galaxies between the LERGs and the HERGs \citep[see reviews by][]{2014ARA&A..52..589H,Tadhunter2016,Hardcastle2020,Magliocchetti2022}. More recent studies, which reach fainter radio luminosities, have found considerable overlap in the accretion rates between the two populations \citep[e.g.][]{Whittam2022}. Here, we consider the accretion rates of the LERGs across the three different galaxy populations, and the HERGs.

The Eddington-scaled accretion rates can be estimated as
\begin{equation}\label{eq:ed_acc_rate}
    \lambda_{\rm{Edd}} = \frac{L_{\rm{rad}} + L_{\rm{mech}}}{L_{\rm{Edd}}},
\end{equation}
where $L_{\rm{rad}}$ is the bolometric radiative luminosity of the AGN, $L_{\rm{mech}}$ is the kinetic/mechanical luminosity from the radio jets, and $L_{\rm{Edd}}$ is the Eddington luminosity limit for each source. In the absence of emission line properties for the vast majority of the sources, the bolometric radiative luminosity is estimated from the best-fit \textsc{cigale} SED model \citep[see][]{Best2023} for each source and is the sum of the AGN accretion disc luminosity and the AGN dust re-emitted luminosity, coming from the re-emitted infrared light due to dust surrounding the AGN. The mechanical luminosity of the jets can be estimated using the 1.4\,GHz radio luminosity as $L_{\rm{kin,\,cav}} = 7 \times 10^{36} f_{\rm{cav}} \left( L_{\rm{1.4\,GHz}} / 10^{25}\, \rm{W\,Hz^{-1}} \right)^{0.68} \rm{W}$ \citep{Cavagnolo2010,2014ARA&A..52..589H}. We translate our observed 150\,MHz radio luminosities to 1.4\,GHz using a spectral index $\alpha = -0.7$. This scaling relation for the mechanical luminosity was determined by using the energy associated with expanding the radio lobes and inflating cavities in the hot X-ray gas observed in massive groups and clusters \citep[e.g.][]{Boehringer1993,Birzan2008,Cavagnolo2010,Timmerman2022}. The Eddington limit for each AGN was then determined using $L_{\rm{Edd}} = 1.31 \times 10^{31} (M_{\rm{BH}}/M_{\odot})\,\rm{W}$, where the black hole mass is estimated using the stellar mass as $M_{\rm{BH}} \approx 0.0014 M_{\star}$ \citep{Haring2004}. Overall, the scaling relations used to estimate the mechanical luminosity and the black hole masses both have a considerable scatter of $\sim$ 0.7\,dex and $\sim$ 0.3\,dex, respectively.

\begin{figure}
    \centering
    \begin{subfigure}{0.5\textwidth}
    \includegraphics[width=0.99\textwidth]{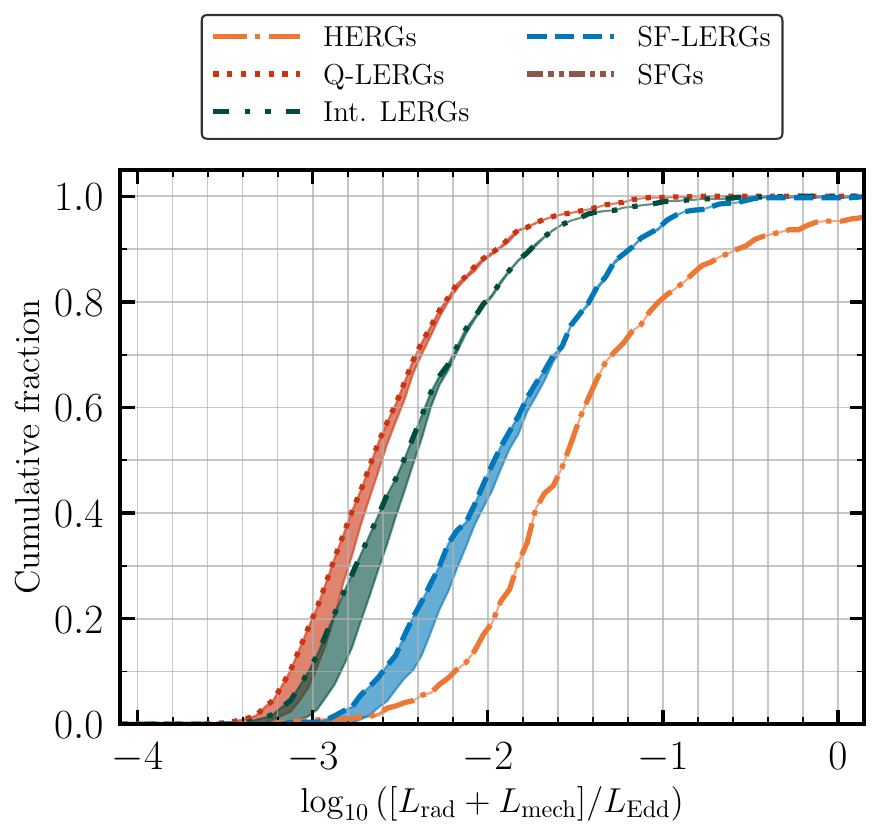}
    \end{subfigure}
    \begin{subfigure}{0.5\textwidth}
    \includegraphics[width=0.99\textwidth]{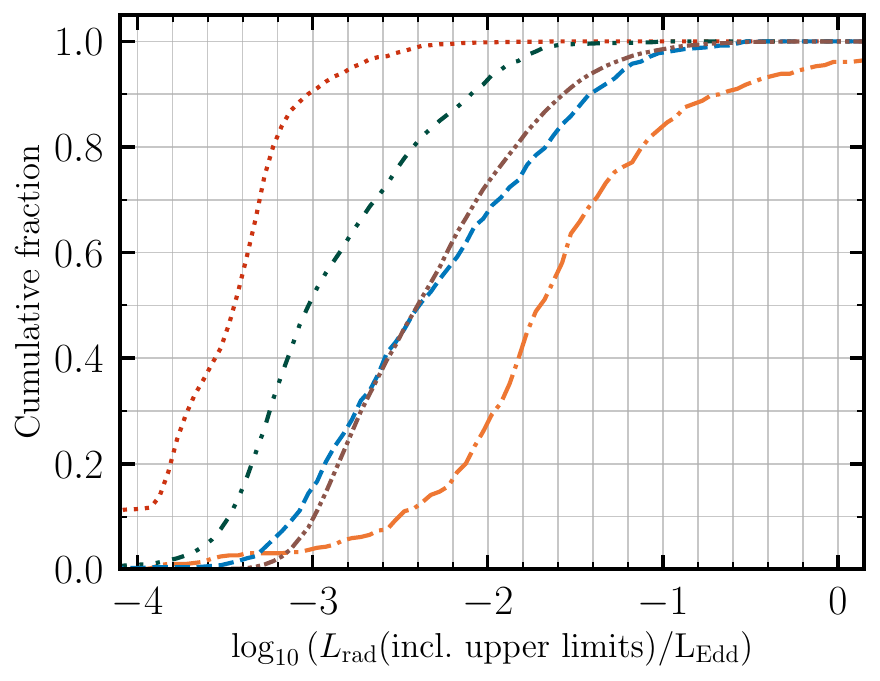}
    \end{subfigure}
    \caption{\label{fig:edd_frac}The cumulative distribution of the Eddington-scaled accretion rates for the LERGs within different galaxy populations, and for the HERGs. The top panel shows the Eddington-scaled accretion rates derived by combining the bolometric radiative AGN luminosity estimated from the SED fitting output and the mechanical jet luminosity. The thick, darker lines correspond to using $L_{\rm{rad}} = 0$ where there are no measurements, and the shaded region represents the effect of setting an upper limit on $L_{\rm{rad}}$ of 5 per cent of the total dust luminosity (see text). The bottom panel shows the maximal accretion rates based on the radiative luminosity only, assigning the 5 per cent upper limit to the sources without measurement.}
\end{figure}

While over 97 per cent of the HERGs have a measured AGN accretion disc and AGN dust luminosity, over 50 per cent of the LERGs hosted by intermediate and star-forming galaxies, and around 80 per cent of the LERGs hosted by quiescent galaxies have no AGN accretion disc or AGN dust luminosity assigned to them. This is in line with expectations as the LERGs are radiatively inefficient AGN and hence not expected to have significant infrared (radiative) luminosities. Sources with extremely low AGN fractions, $f_{\rm{AGN}} < 0.05$, are not assigned a radiative luminosity by \textsc{cigale}; such sources with no measured radiative luminosities, which are predominantly LERGs, can bias our interpretations. To account for this, we derive an upper limit on the bolometric luminosity to be 5 per cent of the total dust luminosity (i.e. corresponding to an $f_{\rm{AGN}} =  0.05$) for sources with no measured bolometric luminosities (see below). 

Using the above relation for the kinetic powers together with Eq.~\ref{eq:ed_acc_rate}, the Eddington-scaled accretion rates were derived for the LERGs and HERGs within the sample, applying the same $L_{\rm{150\,MHz}} \geq 10^{24}\,\rm{W\,Hz^{-1}}$ limit to allow comparisons with the results in the rest of the paper. Fig.~\ref{fig:edd_frac} shows the cumulative accretion rate distributions, where the top panel shows the Eddington-scaled accretion rates determined by combining both the radiative bolometric luminosity and the mechanical luminosity, whereas the bottom panel shows those determined from the radiative bolometric luminosity alone. In the top panel of Fig.~\ref{fig:edd_frac}, the thick darker lines correspond to the accretion rates derived by setting $L_{\rm{rad}} = 0$ for sources with no measured value, while the shaded region indicates the effect of setting these at the 5 per cent upper limit instead, which shifts the distribution to higher values. The Eddington-scaled accretion rates are shown for the HERGs (orange) and the LERGs; the LERGs are further split into those hosted by quiescent (red), intermediate (green), and star-forming (blue) galaxies as previously defined in Sect.~\ref{sec:lerg_sfr_ms_off}.

The HERGs typically have high accretion rates, with $\gtrsim$ 80 per cent above $\lambda_{\rm{Edd}} \sim 0.01$; this is broadly consistent with HERGs being typically fuelled by radiatively-efficient accretion. The quiescent LERGs typically have accretion rates below 1 per cent of the Eddington rate, which is consistent with fuelling occurring in a radiatively inefficient manner from cooling hot gas within their massive host galaxies. The intermediate LERGs show higher accretion rates while still being typically below 1 per cent of the Eddington rate. The star-forming LERGs however have higher accretion rates which do not appear to be consistent with either the HERGs or the other sub-types of LERGs. The bottom panel of Fig.~\ref{fig:edd_frac} shows the distribution considering the bolometric luminosities alone ($L_{\rm{rad}}/L_{\rm{Edd}}$), where the curve is derived using upper limits on $L_{\rm{rad}}$ where there is no measured value, as described above. Through a comparison with the top panel, we find that the radiative bolometric luminosities for the quiescent LERGs are significantly lower than the mechanical jet luminosities, with $L_{\rm{rad}}/L_{\rm{Edd}}$ typically below 0.1 per cent. In contrast, the bolometric radiative luminosities typically dominate over the mechanical jet luminosities for the HERGs. While $\sim$ 70 per cent of the star-forming LERGs have $L_{\rm{rad}}/L_{\rm{Edd}} < 0.01$, the population occupies intermediate values of $L_{\rm{rad}}/L_{\rm{Edd}}$, between the Q-LERGs and HERGs. Based on observations in the local Universe \citep[e.g.][]{2012MNRAS.421.1569B,Mingo2014}, we do not expect a considerable population of LERGs to have accretion rates above 1 per cent of Eddington; we discuss the potential causes for this in Sects~\ref{sec:sclass_edd} and~\ref{sec:lbol_rel}.

\subsection{Potential source classification uncertainties}\label{sec:sclass_edd}
It is possible that source mis-classification may have an impact on the observed accretion rate distributions seen in Fig.~\ref{fig:edd_frac}. In the presence of significant ongoing star-formation, SED fitting routines may be less reliable in determining the presence of low-level AGN emission or robustly separating emission from AGN and star-formation. This could lead to potential mis-classification of sources, in particular between SF-LERGs and HERGs. However, it is important to note that the \citet{Best2023} classification scheme adopted mitigates this issue to an extent in various ways. Firstly, instead of the best-fit $f_{\rm{AGN}}$, they used the 16th percentile of the $f_{\rm{AGN}}$ to identify the presence of radiative-mode AGN (for \textsc{cigale}, the criterion $f_{\rm{AGN, 16}} > 0.06$ was used; $> 0.1$ in Bo\"{o}tes\footnote{A different threshold was used by \citet{Best2023} in Bo\"{o}tes as the $f_{\rm{AGN}}$ values derived from SED fitting were systematically higher, likely due to the differences in how the multi-wavelength catalogue was constructed the Bo\"{o}tes field (see \citealt{Best2023} for full details).}) which is more robust to large uncertainties on the fitted AGN components due to the presence of star-formation. Secondly, in addition to the above, \citet{Best2023} also compared the $\chi^{2}$ values of \textsc{bagpipes} and \textsc{magphys} (which do not model AGN emission) with the values from \textsc{agnfitter} and \textsc{cigale} to identify radiative-mode AGN. \citet{Best2023} validated AGN identified via this classification method with mid-infrared colour-colour diagnostics, finding good agreement (see fig.~5 of \citealt{Best2023}).

We further validate the classification of sources using both recent spectroscopic observations and X-ray data for a subset of the sources, which is described in detail in Appendix~\ref{ap:ap_sclass}. Firstly, we used observations from the Dark Energy Spectroscopic Instrument Early Data Release (DESI-EDR; \citealt{DESICollaboration2023}) to generate stacked optical spectra for each of Q-LERGs, SF-LERGs, HERGs, and SFGs, separately, as classified by \citet{Best2023}. The SF-LERG stack shows strong [O II] emission, indicative of star-formation, and a lower [O III]/[O II] ratio than the HERGs showing that the SF-LERGs are dominated by `normal' SFGs. However, the SF-LERG stack also shows slightly higher [O III] emission than the SFG stack indicating that there may be a small level of radiative-mode AGN contamination. Secondly, we used the \textit{Chandra} Deep Wide-Field Survey (CDWFS; \citealt{Masini2020}) in the Bo\"{o}tes field to study the average X-ray properties of the different source classes. We find that the average (stacked) X-ray luminosities of each of Q-LERGs, intermediate LERGs, and SF-LERGs, show similar levels of X-ray emission, which are considerably lower than the HERGs. Our additional validation presented in Appendix~\ref{ap:ap_sclass} is focused on the average properties of each population rather than a source-by-source comparison, therefore, on average, we conclude that the LERGs hosted by intermediate and star-forming galaxies appear to be distinct to the HERGs, and show similar nuclear properties to the Q-LERGs. While there may be some potential mis-classifications due to the nature of the limitations of the SED-fitting process, these results provide additional confidence that this does not have a significant impact on the robustness of the classifications used in this study.

\subsection{Unreliable $L_{\rm{bol}}$ measurements for SF-LERGs}\label{sec:lbol_rel}
For sources with significant ongoing star-formation, as in the case of the sources identified as SF-LERGs, it is possible that SED fitting routines may not be able to reliably model the contribution from AGN and star-formation simultaneously, particularly for fainter galaxies with fewer robust photometric data points. To investigate this, we derive an equivalent $L_{\rm{rad}}/L_{\rm{Edd}}$ distribution for sources classified as SFGs\footnote{Here, SFGs are defined as sources that do not display signs of either radio-excess AGN or radiative-mode AGN.} by \citet{Best2023} that also meet the star-forming criteria based on the the main-sequence defined in Sect.~\ref{sec:lerg_sfr_ms_off}; this is shown by the brown curve in Fig.~\ref{fig:edd_frac} (bottom panel). We find that  $>$ 60 per cent of the SFGs have no measured AGN fraction ($f_{\rm{AGN}}$; i.e. no measured `radiative' luminosities). The remaining sources have very low $f_{\rm{AGN}}$ values (typically $\lesssim$ 0.05), but given the high total luminosities associated with ongoing star-formation within the galaxies this still results in non-negligible $L_{\rm{rad}}$. Hence, the $L_{\rm{rad}}/L_{\rm{Edd}}$ distribution for the SFGs appears at intermediate values, and indeed is very similar to the SF-LERG distribution when incorporating the limits. In comparison, $\sim$ 90 per cent of the SF-LERGs have $f_{\rm{AGN}} \leq 0.05$, indicating low contribution from AGN emission. The similarities in the distributions of the two populations, where the key difference between how the two populations were selected is the presence of a radio-excess AGN for the SF-LERGs, highlights that the higher $L_{\rm{rad}}/L_{\rm{Edd}}$ observed for the SF-LERGs compared to Q-LERGs is primarily due to wrongly-assigned luminosity from star-formation processes rather than (missed) significant AGN emission arising from mis-classified HERGs. Overall, this result suggests that the radiative luminosities estimated for the SF-LERGs through SED fitting processes may not be robust.

\subsection{Comparison of Eddington-scaled accretion rates with literature}
We now compare our results on the Eddington-scaled accretion rates of LERGs and HERGs shown in Fig.~\ref{fig:edd_frac} with other studies in the literature. Following the above caveats regarding the robustness of the radiative luminosities estimated from SED fitting for the SF-LERGs, we do not attempt to over-interpret the Eddington-scaled accretion rates derived for this sub-group of LERGs in Fig.~\ref{fig:edd_frac}. Instead, here we primarily focus on comparing our Q-LERG and the HERG results to previous works, where we expect the derived radiative luminosities to be most robust.

Our results for the HERGs and the \textit{quiescent} LERGs are in agreement with previous studies of the HERG and \textit{overall} LERG populations in the local Universe \citep[e.g.][]{2012MNRAS.421.1569B,Mingo2014}. We find that there is more overlap between the LERG and HERG populations in our study, and while it is possible that this is a real effect, we suggest that these observations are also consistent with the combination of measurement uncertainties and selection effects. In particular, aside from the robustness of the radiative luminosities for SF-LERGs discussed above, our use of a scaling relationship to estimate black hole masses has considerable scatter, compared to the more robust measurements by \citet{2012MNRAS.421.1569B} using spectra. Such scatter in the empirical scaling relationship can spread out the intrinsic Eddington-rate distributions of the two populations resulting in more apparent overlap; prospects for overcoming these limitations are discussed in Sect.~\ref{sec:fuelling_discussion} and~\ref{sec:conclusions}. Moreover, as we showed in Fig.~\ref{fig:lerg_frac_sfr_ms_mass} (and in \citealt{Kondapally2022}), SF-LERGs are expected to be less common at low redshifts studied by \citet{2012MNRAS.421.1569B}. Additionally, the use of emission line diagnostics to select LERGs by \citet{2012MNRAS.421.1569B}, which requires low line fluxes or equivalent widths from the [O II] and [O III] lines may select against star-forming hosts of LERGs.

Recently, \citet{Whittam2022} used MIGHTEE early-science observations of the COSMOS field at 1.4\,GHz to study the properties of LERGs and HERGs. Their data covers an area of $\sim 1\,\rm{deg^{2}}$ and reaches a noise level of $\sim 4\,\mu \rm{Jy/beam}$, comparable in depth to our LoTSS Deep Fields data assuming a standard spectral index. \citet{Whittam2022} found considerable overlap in the Eddington-scaled accretion rates between their LERGs and HERGs. Overall, \citet{Whittam2022} have suggested that at higher redshifts and lower radio luminosities, the distinction between the LERG and HERG populations becomes less clear, with their host galaxy properties also becoming more similar. However, \citet{Whittam2022} also use the same scaling relationships to determine the jet mechanical luminosity and the black hole masses as our study, which as discussed above leads to a spread in the derived Eddington-scaled accretion rate distributions, and hence more overlap in the LERG and HERG populations. It is worth noting that the classification of LERGs at higher redshifts by \citet{Whittam2022} are less robust, in particular due to the X-ray data which limits their X-ray AGN classification to $z < 0.5$. As a result, they are only able to classify sources as `LERGs' at $z < 0.5$, and instead identify a sample of `probable LERGs' at higher redshifts. The Eddington-rate distribution for their `probable LERG' sample extends to higher values than their LERG sample; it is therefore possible that their `probable LERGs' could be contaminated by mis-classified HERGs, and could suffer the same potential biases as our SF-LERGs whereby their radiative luminosities, and hence accretion rates, may be over-estimated due to the presence of strong star-formation. If this is not the case, then this would imply that the SF-LERGs indeed have higher bolometric luminosities, which would have interesting implications for the nature of these sources. We also note that the sample used by \citet{Whittam2022} probes higher redshifts than our sample, which may also impact the detailed comparison with their study.

\section{On the fuelling of LERGs and HERGs across the galaxy population}\label{sec:fuelling_discussion}
The dependence of AGN activity on stellar mass and star-formation rate presented in Sects.~\ref{sec:flerg_fherg_mass} and~\ref{sec:lerg_sfr_ms_off} can be combined with the accretion rate properties derived in Sect.~\ref{sec:edd_dist} to understand the nature of the faint LERG and HERG populations in the early Universe as traced by our observations. Our results from Fig.~\ref{fig:lerg_frac_sfr_ms} and Fig.~\ref{fig:lerg_frac_sfr_ms_mass} suggest that LERGs may be fuelled by multiple processes based on the available fuel supply and the stellar mass. Fuelling from hot gas is more likely to occur in massive galaxies, which also host massive black holes, and such fuelling typically occurs at low accretion rates and leads to the formation of a LERG. Our observations of LERGs across the galaxy population support a scenario where this process likely fuels the AGN in massive galaxies regardless of their star-formation activity, effectively setting a `minimum' level of LERG activity. This naturally explains the existence of the quiescent LERGs, which have low accretion rates (as seen in Fig.~\ref{fig:edd_frac}) that can be associated with fuelling from hot gas, and which are found in massive galaxies, resulting in the steep stellar mass relation observed in Fig.~\ref{fig:lerg_frac_sfr_ms}. In massive galaxies with higher star-formation rates, the same process may also fuel the AGN; this is supported by our results in Fig.~\ref{fig:lerg_frac_sfr_ms_mass} which show that at high masses, all galaxies, regardless of galaxy type, have roughly the same level of LERG activity (or a marginally higher level in star-forming LERGs), pointing to a common fuelling process from hot gas. We note that the similarities in the incidence of LERGs in massive galaxies, which have massive dark matter haloes, may also suggest that the environment is an important factor in accretion from hot gas \citep[see e.g.][]{Tasse2008,Sabater2013,Mingo2019,Hardcastle2020}; a detailed modelling of the environments of these AGN is outside the scope of this paper.

In galaxies with high levels of star-formation activity, there is an abundant supply of cold gas which can fuel both star-formation and black hole growth. Therefore, we suggest that there may be an additional fuelling mechanism associated with the cold gas that fuels LERGs in galaxies with higher star-formation activity; this is particularly important in lower mass galaxies. In this scenario, we would expect a higher level of LERG activity in galaxies with higher star-formation activity at a fixed stellar mass; we observe a factor of 2--10 higher LERG incidence in star-forming galaxies compared to quiescent galaxies, as seen in Fig.~\ref{fig:lerg_frac_sfr_ms_mass}. Additionally, in both Fig.~\ref{fig:lerg_frac_sfr_ms} and Fig.~\ref{fig:lerg_frac_sfr_ms_mass}, we see that for a given stellar mass, the incidence of LERGs in star-forming galaxies increases with increasing redshift, when the gas fractions were also higher. Moreover, \citet{Heckman2024} find that the evolution of the cosmic energy budget from AGN jets is well correlated with the quenching rate, suggesting that the SF-LERGs may be connected to the quenching of star-forming galaxies. We note that although the presence of abundant cold gas is capable of fuelling the AGN at high rates, this does not necessarily mean that this always occurs \citep[see][]{Hardcastle2018}; therefore, if LERGs are defined as AGN fuelling below 1 per cent of the Eddington-rate, then it is possible for cold gas to supply this. We note that the stochastic nature of accretion can be coupled to the different timescales involved in the generation of the radio jets and accretion disc structure, which can complicate the interpretations.

The picture for the HERGs appears more straightforward, with HERGs typically being found in star-forming systems, where they are fuelled at high accretion rates in a radiatively efficient manner by the cold gas present within their star-forming host galaxies, broadly consistent with previous studies at lower redshift \citep[e.g.][]{2012MNRAS.421.1569B,Mingo2014}. It is not clear how the population of HERGs accreting at low rates fit into this picture, however we note that the large scatter in the black hole mass - stellar mass relationship, and in the radio luminosity - jet kinetic power relationship may broaden the observed accretion rate distribution.

\section{Conclusions}\label{sec:conclusions}
In this paper, we have used data from the LOFAR Deep Fields to study the host galaxy properties of radio-AGN out to $z = 1.5$. The three fields studied, ELAIS-N1, Lockman Hole, and Bo\"{o}tes, all benefit from having deep wide-area coverage from the UV to far-infrared wavelengths over $\sim 25\,\rm{deg^{2}}$. This dataset was used to measure robust physical galaxy properties and for the identification of a sample of radio-AGN, which we further split into LERGs and HERGs. The host galaxy properties of these AGN were compared to a mid-infrared selected parent sample of the underlying galaxy population. Using this, we have presented robust measurements of the incidence of radio-AGN activity on stellar mass, star-formation rate, and black hole accretion rate across $0.3 < z \leq 1.5$. We divided the AGN population based on their star-formation activity relative to the main-sequence of star-formation into three groups: quiescent, intermediate, and star-forming hosts. In our analysis, we compute the Eddington-scaled accretion rates, $\lambda_{\rm{Edd}}$ for the AGN, by combining the bolometric radiative luminosity and mechanical luminosity from the jets. Our main conclusions are as follows:

\begin{enumerate}
    \item The LERGs show a steep dependence on stellar mass with some evidence of flattening at higher redshifts, as found by previous studies. The HERGs show a similarly steep relation, but with significantly lower prevalence than the LERGs at all stellar masses.

    \item HERGs are predominantly found in galaxies with ongoing star-formation activity, across redshift, suggesting the need for cold gas supply to trigger HERG activity. In contrast, we find that LERG activity occurs across the galaxy population, in both quiescent and star-forming galaxies, with LERGs predominantly hosted by star-forming galaxies in the early Universe.

    \item The HERGs typically accrete above 1 per cent of the Eddington-scaled accretion rates, which is broadly consistent with previous studies in the local Universe.
    
    \item When split into different galaxy groups based on star-formation activity, the quiescent LERGs show typical accretion rates below 1 per cent of the Eddington-rate; this is consistent with results for the \textit{overall} LERG population in the local Universe. We find that the radiative luminosities (and hence accretion rates) derived for SF-LERGs from SED fitting are likely not robust; this highlights the need for spectroscopic data to robustly characterise AGN accretion properties.

    \item The incidence of LERGs in quiescent galaxies has a steep dependence on stellar mass which does not vary with redshift; this is found to be consistent with fuelling occurring at low Eddington rates from hot gas. 
    
    \item LERG activity depends on both stellar mass and star-formation rate, suggesting a different fuelling mechanism of LERGs across the galaxy population. We find that at high masses, LERGs are roughly equally as likely to be found in quiescent or star-forming galaxies: at high masses, $\sim$ 10--20 per cent of the galaxies host a LERG irrespective of the star-formation activity.
    
    \item At lower masses, LERG activity is significantly more likely to be hosted in galaxies with higher star-formation activity, by a factor of up to 10 compared to quiescent galaxies. This suggests that at lower masses, in particular within star-forming galaxies, there may be an additional mechanism that triggers LERG activity.
\end{enumerate} 

In summary, our recent work has found a significant population of LERGs hosted by star-forming galaxies, in contrast to expectations from studies in the local Universe. Our observations suggest that accretion from the hot gas provides a minimum level of incidence for the LERGs with an additional fuelling mechanism from cold gas providing an enhancement. In massive galaxies where the incidence is high, hot gas accretion dominates the fuelling across the galaxy population, regardless of star-formation activity. However in low mass galaxies, fuelling from cold gas provides an enhancement of LERG activity within star-forming galaxies in particular. The precise fuelling mechanisms that trigger LERG activity in star-forming galaxies remain unclear at present, although our results suggest that this may be occurring from cold gas. Further evidence in support of this scenario comes from the incidence of LERGs increasing with both increasing star-formation activity and at higher redshifts, when the gas fractions were also higher.

Going beyond the global galaxy properties towards a characterisation of their molecular gas properties, in comparison to a matched sample of LERGs and HERGs in different galaxy types, will be necessary to understand the fuelling processes at play. Our results also find some overlap in the Eddington-scaled accretion rates of LERGs and HERGs; it is possible that some of this observed overlap could be associated with the stochastic nature of accretion, scatter in the empirical scaling relations, or potential uncertainties in source classifications and derived radiative luminosities. To overcome remaining uncertainties in source classification and in estimating galaxy and AGN properties requires the availability of spectra; emission line diagnostics allows the most robust method of identifying radiatively efficient versus inefficient activity \citep[e.g.][]{2012MNRAS.421.1569B,Whittam2018}. A similar analysis has recently been applied to the wide-area LoTSS DR2 dataset (\citealt{Hardcastle2023,Drake2024}) using SDSS spectroscopic observations; however current spectroscopic datasets are only able to cover a small fraction of the LOFAR-detected sources. The upcoming WEAVE-LOFAR survey \citep{2016sf2a.conf..271S}, using the new WEAVE instrument \citep{Jin2024} on-board the William Herschel Telescope, will provide dedicated follow-up optical spectra for a large number of LOFAR-detected sources. This will enable robust source classification and identification of different AGN types, and the characterisation of their properties including radiative luminosities and black hole masses.

\section*{Acknowledgements}

This paper is based (in part) on data obtained with the International LOFAR Telescope (ILT) under project codes LC0\_015, LC2\_024, LC2\_038, LC3\_008, LC4\_008, LC4\_034 and LT10\_01. LOFAR \citep{2013A&A...556A...2V} is the Low Frequency Array designed and constructed by ASTRON. It has observing, data processing, and data storage facilities in several countries, which are owned by various parties (each with their own funding sources), and which are collectively operated by the ILT foundation under a joint scientific policy. The ILT resources have benefitted from the following recent major funding sources: CNRS-INSU, Observatoire de Paris and Université d'Orléans, France; BMBF, MIWF-NRW, MPG, Germany; Science Foundation Ireland (SFI), Department of Business, Enterprise and Innovation (DBEI), Ireland; NWO, The Netherlands; The Science and Technology Facilities Council, UK; Ministry of Science and Higher Education, Poland. For the purpose of open access, the author has applied a Creative Commons Attribution (CC BY) licence to any Author Accepted Manuscript version arising from this submission. This research made use of {\sc Astropy}, a community-developed core Python package for astronomy \citep{astropy:2013, astropy:2018} hosted at \url{http://www.astropy.org/}, and of {\sc Matplotlib} \citep{hunter2007matplotlib}.

RK and PNB acknowledge support from the UK Science and Technologies Facilities Council (STFC) via grant ST/V000594/1. MIA acknowledges support from the UK Science and Technology Facilities Council (STFC) studentship under the grant ST/V506709/1. KM has been supported by the Polish National Science Centre (UMO-2018/30/E/ST9/00082). BM acknowledges support from the UK STFC under grant ST/T000295/1. BM further acknowledges support from UKRI STFC for an Ernest Rutherford Fellowship [grant number ST/Z510257/1]. SD acknowledges support from a Science and Technology Facilities Council (STFC) studentship via grant ST/W507490/1. IP acknowledges support from INAF under the Large Grant 2022 funding scheme (project ``MeerKAT and LOFAR Team up: a Unique Radio Window on Galaxy/AGN co-Evolution''). DJBS acknowledges support from the UK Science and Technology Facilities Council (STFC) under grants ST/V000624/1 and ST/Y001028/1. We thank Dr. James Aird for helpful discussions on the X-ray stacking analysis. For the purpose of open access, the author has applied a Creative Commons Attribution (CC BY) licence to any Author Accepted Manuscript version arising from this submission.

\section*{Data Availability}
The dataset used in this study comes primarily from the LoTSS Deep Fields Data Release 1. The corresponding radio data are presented by \citet{Tasse2021,Sabater2021}, the multi-wavelength data, host-galaxy counterparts, and photometric redshifts are presented by \citet{Kondapally2021,Duncan2021}, with the source classifications presented by \citet{Best2023}. The images and catalogues are publicly available at \url{https://lofar-surveys.org/deepfields.html}. The SED fitting output for the mid-infrared parent sample is presented by \citet{Smith2021}.



\bibliographystyle{mnras}
\bibliography{lofar_deepfields_agn} 




\appendix

\section{Radio-excess selection completeness corrections}\label{ap:re_comp}
\begin{figure*}
    \centering
    \begin{subfigure}{0.5\textwidth}
    \centering
    \includegraphics[width=\textwidth]{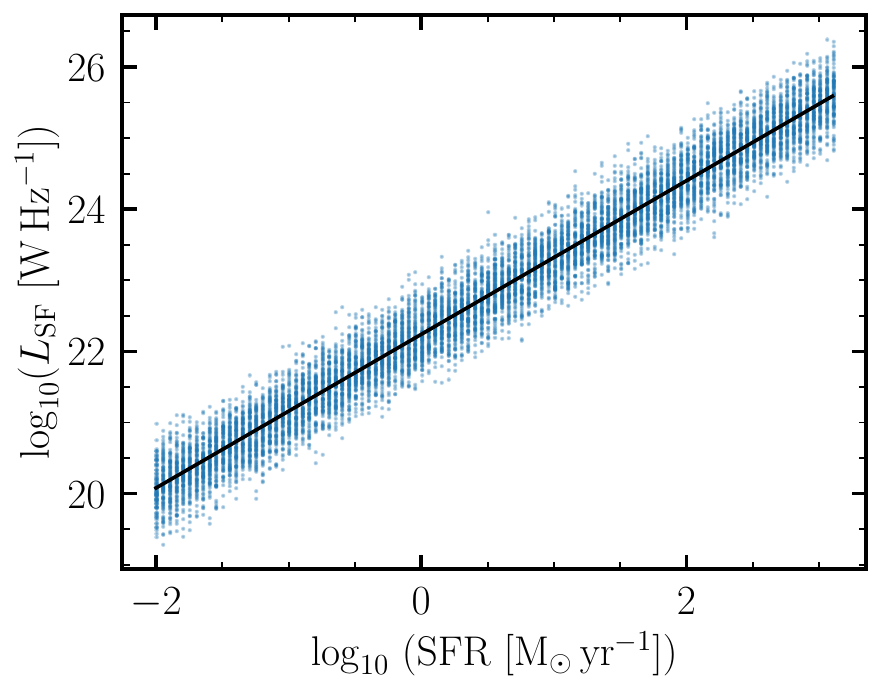}
    \end{subfigure}%
    \begin{subfigure}{0.5\textwidth}
    \includegraphics[width=0.983\textwidth]{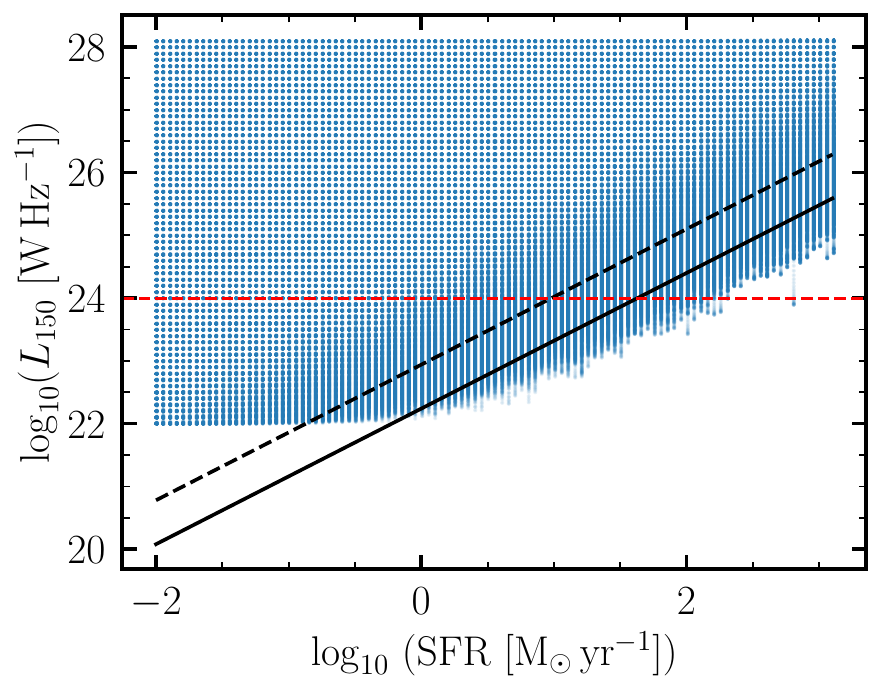}
    \end{subfigure}
    \caption{\label{fig_ap:sfr_l150_sim}Results of the Monte Carlo simulation to quantify the completeness of the radio-excess AGN selection. \textit{Left:} The simulated relationship between the 150\,MHz radio-luminosity and star-formation rate for star-forming galaxies, with a simulated Gaussian scatter of 0.3\,dex. \textit{Right:} The total 150\,MHz radio luminosity for simulated sources, calculated as the sum of the $L_{\rm{AGN}}$ and $L_{\rm{SF}}$ components (from the left panel), as a function of the star-formation rate (see text). The black solid line in both panels shows the radio luminosity -- star-formation rate relation of \citet{Best2023}. The dashed black line shows the radio-excess criterion (0.7\,dex above the relation) used to identify radio-excess AGN. The red line corresponds to the radio luminosity limit of $L_{\rm{150\,MHz}} \geq 10^{24}\,\rm{W\,Hz^{-1}}$ applied for the analysis }
\end{figure*}

\begin{figure}
    \centering
    \includegraphics[width=\columnwidth]{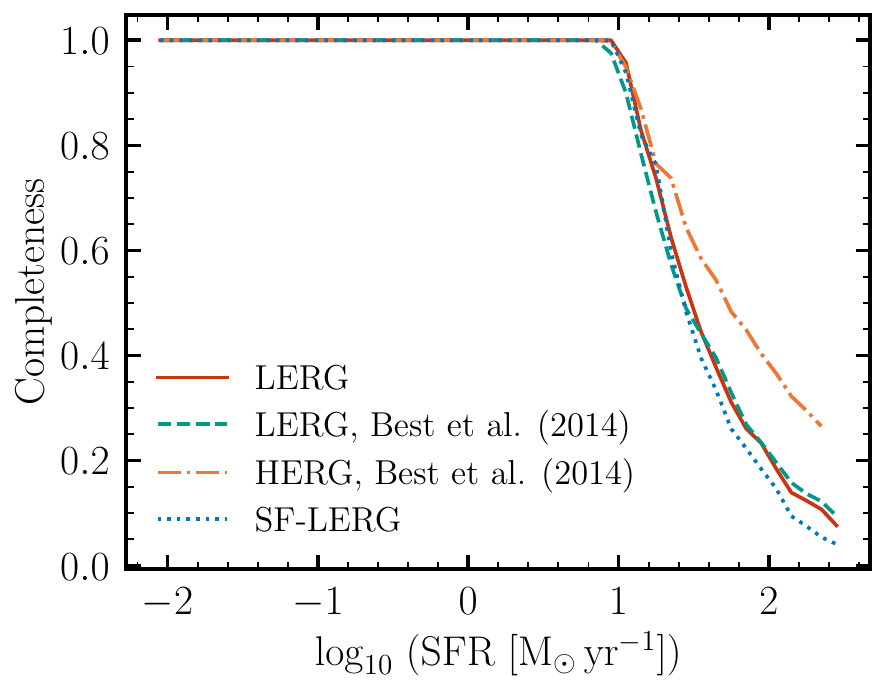}
    \caption{\label{fig_ap:re_sfr_comp}Radio-excess AGN selection completeness as a function of SFR for LERGs, SF-LERGs, and HERGs. The completeness curves are derived by convolving the Monte Carlo simulation with the radio luminosity function (see text). Therefore, in each case, the curves correspond to the completeness in the radio-excess selection for the specific AGN type above a limiting luminosity of $L_{\rm{150\,MHz}} \geq 10^{24}\,\rm{W\,Hz^{-1}}$.}
\end{figure}

The selection of AGN using the radio-excess based on the radio-luminosity -- SFR relation may lead to mis-classification of low (radio) luminosity AGN as SFGs. Consider that the 150\,MHz radio luminosity for each source is the sum of the radio luminosity arising from the AGN (jets) and from star-formation processes as follows, $L_{\rm{150\,MHz}} = L_{\rm{AGN}} + L_{\rm{SF}}$. Therefore, applying a $L_{\rm{150\,MHz}}$ limit, as done throughout most of the analysis in this paper, will not uniformly select AGN above the same luminosity limit in all galaxies. Specifically, for a given AGN luminosity, sources hosted in more star-forming galaxies will be less likely to satisfy the radio-excess criterion, compared to quiescent host-galaxies. This introduces an incompleteness in the radio-excess AGN selection at high star-formation rates. This is particularly relevant for our analysis, where we apply a 150\,MHz radio luminosity limit ($L_{\rm{150\,MHz}} \geq 10^{24}\,\rm{W\,Hz^{-1}}$) to account for the radio flux-density incompleteness effects out to $z \sim 1.5$ (see Sec~\ref{sec:flerg_fherg_mass}). Choosing a sufficiently high radio luminosity limit would ensure a largely complete sample of luminous AGN, however would limit the analysis to the most luminous sources and results in poorer statistics. 

An alternative approach is to assess the impact of the above selection effect on the results derived in this paper, which is described here. We first generated mock sources by simulating their radio luminosities, which are assumed to be described by a two-component model, $L_{\rm{150\,MHz}} = L_{\rm{AGN}} + L_{\rm{SF}}$, where $L_{\rm{AGN}}$ is the AGN component, and $L_{\rm{SF}}$ is the star-formation component to the radio luminosity. There is a tight correlation between the radio luminosity and SFR of star-forming galaxies \citep[e.g.][]{gurkan2018lofar_sfr,Smith2021}. The $L_{\rm{SF}}$ component was modelled as a Gaussian centred on the radio luminosity -- SFR relation derived for the LoTSS-Deep sample by \citet{Best2023} and used for the radio-excess selection, $\log_{10}(L_{\rm{SF, mean}}/\rm{W\,Hz^{-1}}) = 22.24 + 1.08 \log_{10}(\rm{SFR/M_{\odot}\,yr^{-1}})$, with a standard deviation of 0.3\,dex (following the typical scatter in the relationship; see \citealt{Smith2021} and \citealt{Cochrane2023}). SFRs spanning 5 orders of magnitude, between $-2 < \log_{10} (\rm{SFR}/\rm{M_{\odot}\,yr^{-1}}) < 3$ were used to calculate $L_{\rm{SF}}$, generating a distribution of 10,000 simulated sources along the $L_{\rm{SF}} - \rm{SFR}$ plane. For visualisation purposes, we show the resulting distribution for a fraction of the sources in Fig.~\ref{fig_ap:sfr_l150_sim}. The $L_{\rm{AGN}}$ component was finely sampled in 100 bins from a uniform distribution spanning $L_{\rm{150\,MHz}} = 10^{23} - 10^{28}\,\rm{W\,Hz^{-1}}$, covering the full range of luminosities probed by the AGN sample used in this analysis \citep[see][]{Kondapally2022,Best2023}. 

For each $L_{\rm{AGN}}$ value, the total 150\,MHz radio luminosity, $L_{\rm{150\,MHz}}$, was calculated by summing the two components together, to transform the above into a distribution of 10,000 sources along the $L_{\rm{150\,MHz}} - \rm{SFR}$ plane. This process is repeated for each $L_{\rm{AGN}}$ to capture the effect of varying the AGN luminosity on the total 150\,MHz luminosity along this plane by generating a sample of $100 \times 10,000$ sources. For visualisation purposes again, we show this distribution for a fraction of the simulated sources in the right panel of Fig.~\ref{fig_ap:sfr_l150_sim}. As the completeness depends on both radio luminosity and SFR, we split this large sample into narrow bins in $L_{\rm{150\,MHz}}$ and SFR, and for each bin, we calculated the fraction of simulated sources (weighted by the radio luminosity function; see below) that would satisfy the radio-excess selection, as in, with $L_{\rm{150\,MHz}}$ above 0.7\,dex from $L_{\rm{SF, mean}}$ (dashed black line in Fig.~\ref{fig_ap:sfr_l150_sim}) to determine the completeness of the radio-excess selection in each bin.

Throughout the analysis in this paper, we have used a radio luminosity limit of $L_{\rm{150\,MHz}} \geq 10^{24}\,\rm{W\,Hz^{-1}}$, and performed the analysis for different AGN populations which have distinct radio luminosity functions. To derive appropriate completeness corrections as a function of star-formation rate, we weight the completeness along the SFR axis by the radio luminosity function, down to $10^{24}\,\rm{W\,Hz^{-1}}$ to derive the radio-excess selection completeness corrections as a function of SFR for different AGN populations, which are shown in Fig.~\ref{fig_ap:re_sfr_comp}. The radio AGN luminosity functions are often modelled as a broken power-law \citep[e.g.][]{1990MNRAS.247...19D}. \citet{Kondapally2022} fit the LERG luminosity function for the same sample that is used in this study, however they compute this over different redshift bins. We use the $0.5 < z \leq 1$ fit to the radio luminosity function from \citet{Kondapally2022} to compute the corrections (red solid line). Their luminosity function was found to be in good agreement with that of \citet{2014MNRAS.445..955B}, and hence the good agreement seen with the corrections determined using the \citet{2014MNRAS.445..955B} LERG luminosity function (green dashed line). \citet{Kondapally2022} were not able to fit a model to the HERG luminosity function as the bright-end slope was largely unconstrained. Therefore, we use the $0.5 < z \leq 1$ best-fit HERG luminosity function from \citet{2014MNRAS.445..955B}, given the good agreement between our measurements and their LERGs (orange dosh dotted line). We find that the LERG luminosity functions do not evolve significantly between $z = 1$ and $z = 1.5$, and as a result, the derived completeness corrections are similar. We also compute a correction for the subset of LERGs hosted by star-forming galaxies (SF-LERGs). Although \citet{Kondapally2022} do not produce a fit to their SF-LERG luminosity functions, we use the difference in their fitted models to the total LERG and quiescent LERG luminosity functions to compute the SF-LERG corrections shown in Fig.~\ref{fig_ap:re_sfr_comp} (blue dotted line).

In summary, we find that for the radio-excess selection criteria, we achieve a near 100 per cent completeness for SFR $<$ 10\,$\rm{M_{\odot}yr^{-1}}$, which can also be seen from the cross-over point in Fig.~\ref{fig_ap:sfr_l150_sim}. At higher star-formation rates, the completeness falls quickly, but we note that the choice of the luminosity functions do not have a significant effect on the derived completeness corrections. 

\section{Characterisation of the effect of source classification uncertainties}\label{ap:ap_sclass}
\begin{figure*}
    \centering
    \begin{subfigure}[b]{0.49\textwidth}
    \includegraphics[width=0.99\linewidth]{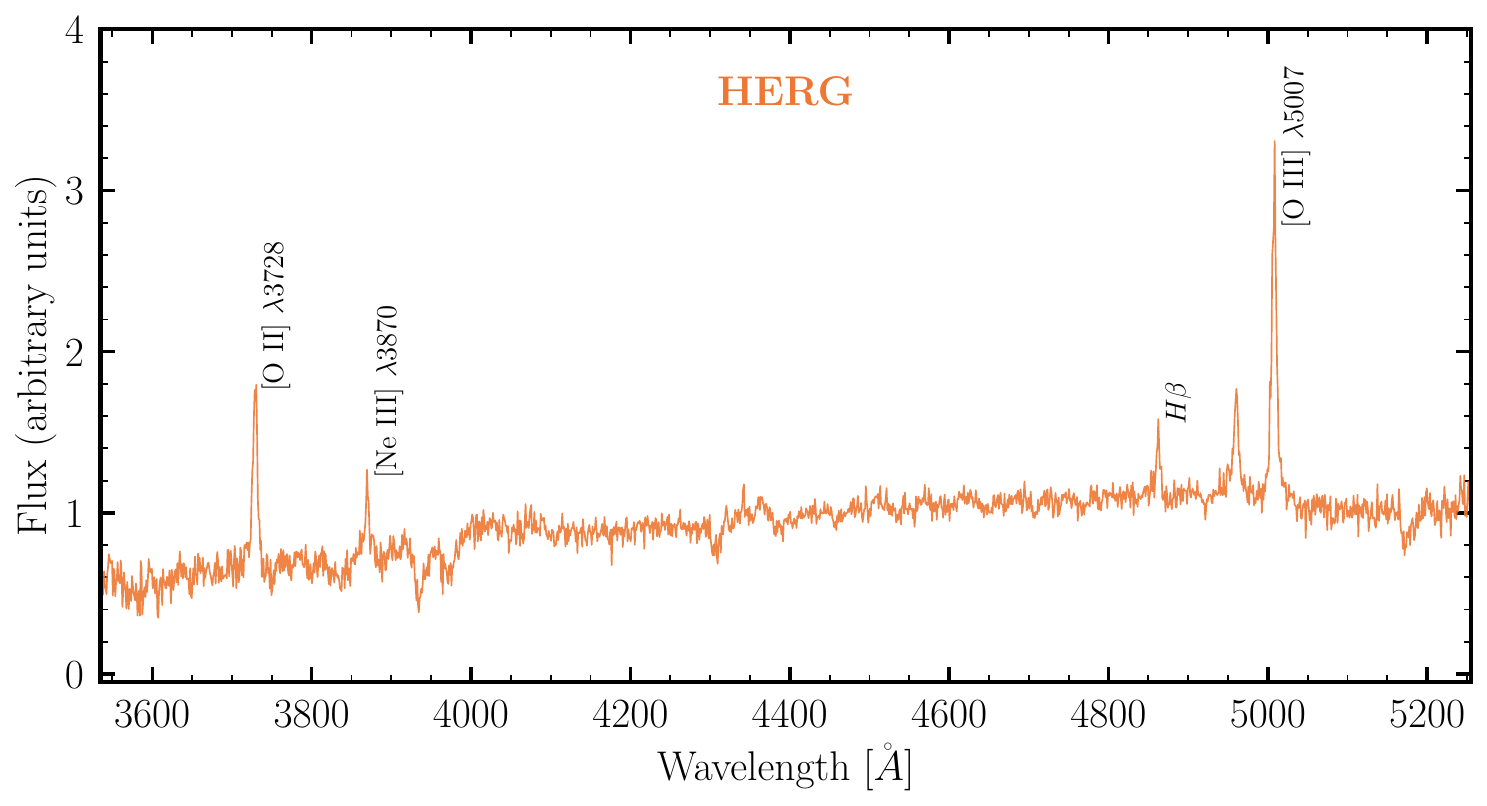}
    \end{subfigure}
    \hfill
    \begin{subfigure}[b]{0.49\textwidth}
    \includegraphics[width=0.99\linewidth]{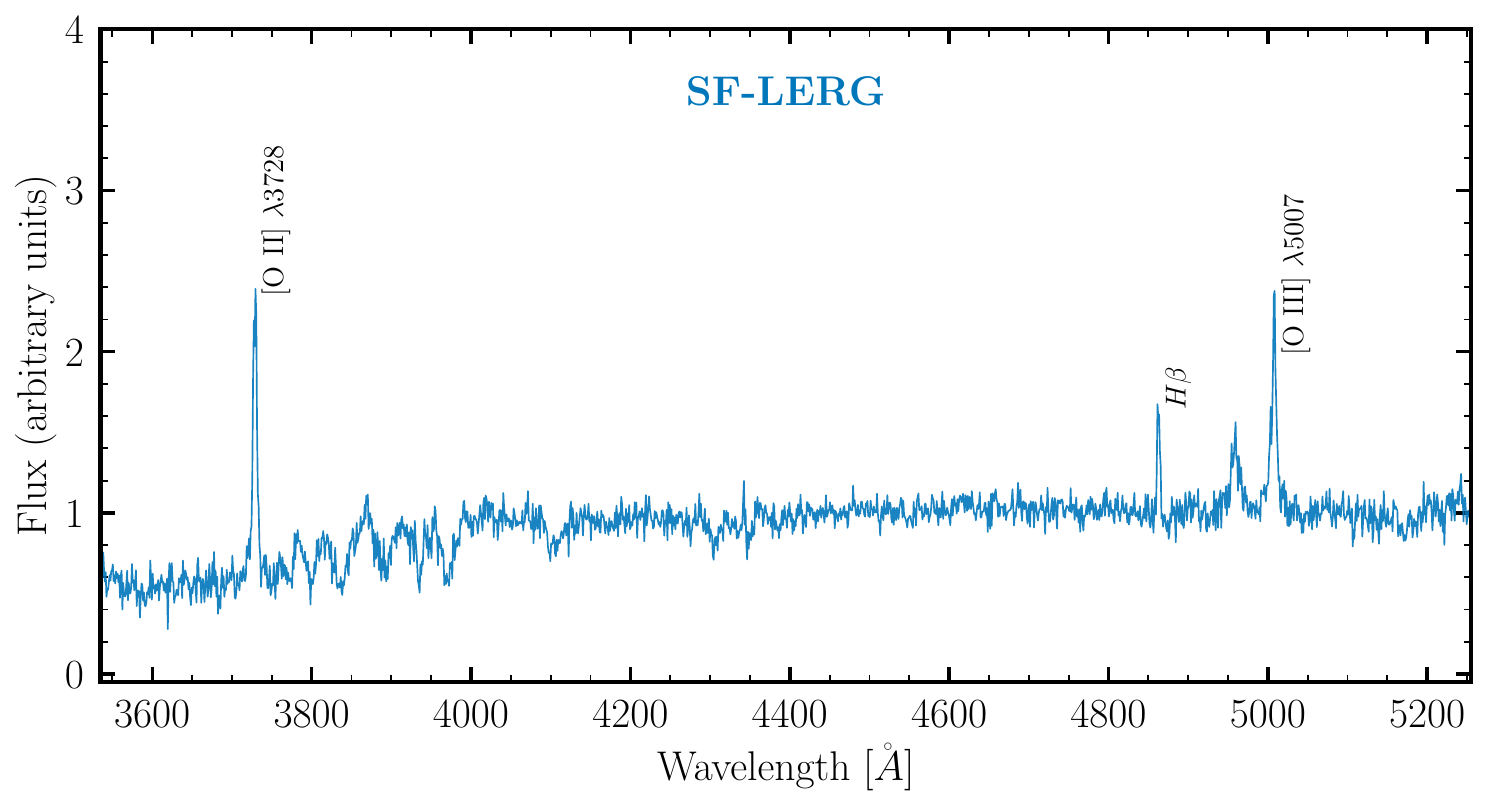}
    \end{subfigure}
    \begin{subfigure}[b]{0.49\textwidth}
    \includegraphics[width=0.99\linewidth]{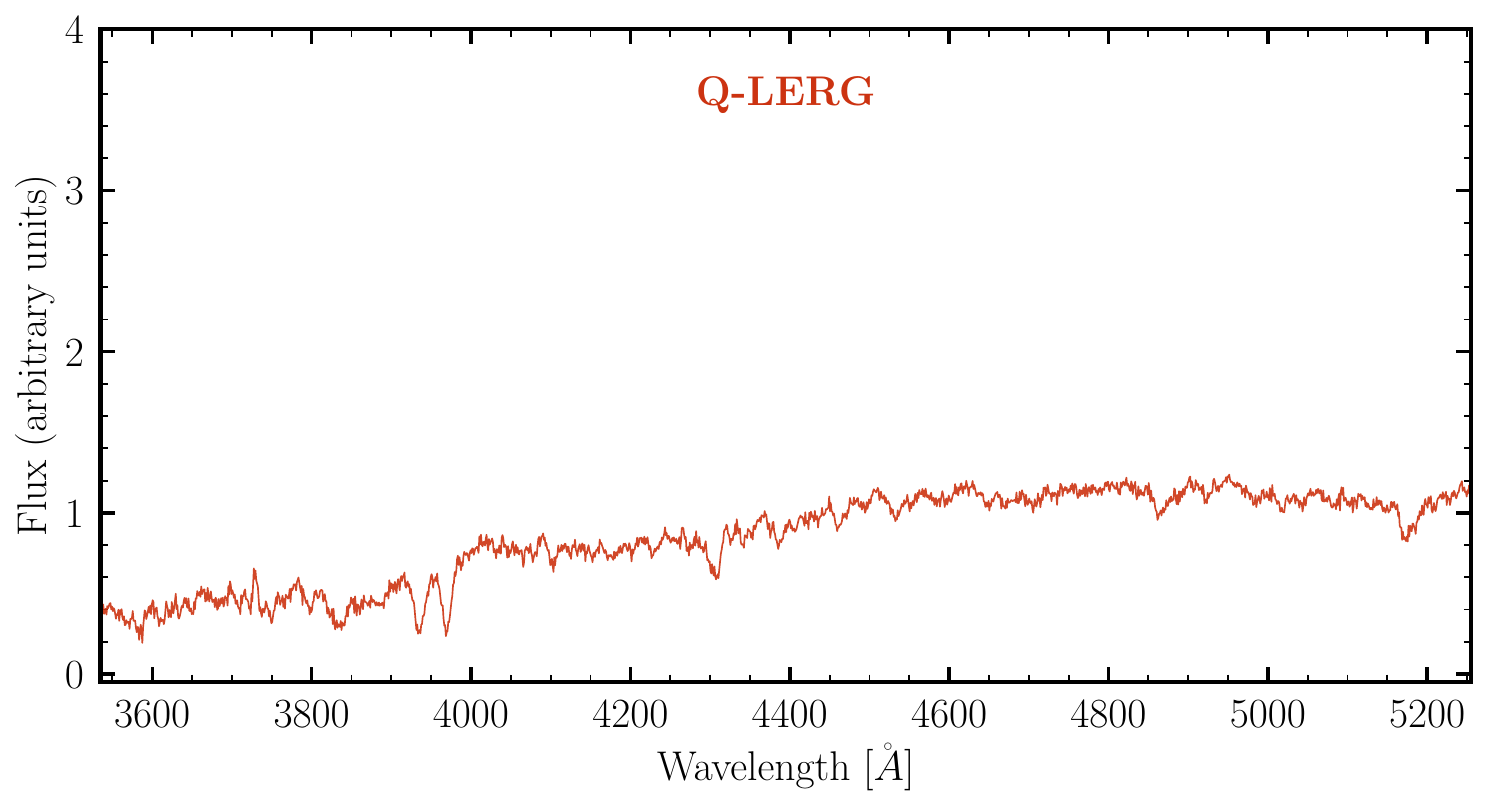}
    \end{subfigure}
    \hfill
    \begin{subfigure}[b]{0.49\textwidth}
    \includegraphics[width=0.99\linewidth]{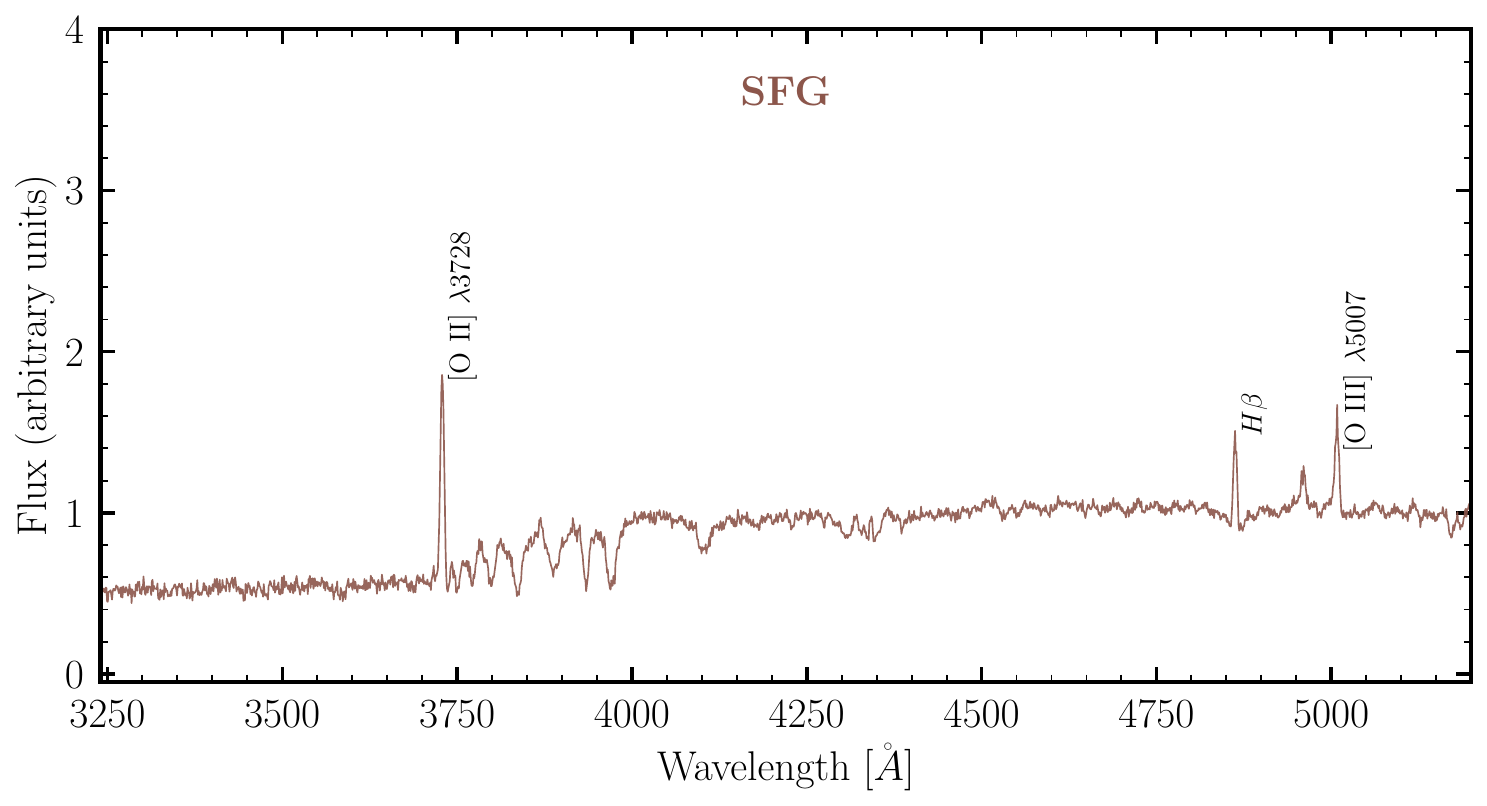}
    \end{subfigure}
    \caption{\label{fig_ap:spec_stack}The median stacked spectra for HERGs, SF-LERGs, Q-LEGRs, and SFGs across $0.5 < z \leq 1$. The individual spectra, derived from the DESI EDR \citep{DESICollaboration2023}, are shifted to the rest-frame, normalised and resampled onto a common wavelength grid before computing a median stacked spectrum. The Q-LERG stack shows Ca H and K absorption features, indicative of an old stellar population, with no signs of emission lines associated with either star-formation or AGN activity. The HERG stack show strong detection of high-excitation emission lines, such as [Ne III] and [O III], and an [O III]/h$\beta$ ratio associated with the presence of AGN activity.}
\end{figure*}

In this section we assess the robustness of the SED-fitting based source classifications by using additional diagnostics that incorporate spectroscopic observations and X-ray observations. 

\subsection{Spectral stacking of LERGs and HERGs}\label{ap:spec_stack}
At the time of the first data release of LoTSS-Deep, the vast majority of the radio sources lacked spectroscopic data, which provide the most robust method of classifying different modes of AGN using emission line diagnostics. As a result, the source classifications used in this study \citep[see][]{Best2023} are based primarily on photometric SED fitting incorporating UV to X-ray datasets in order to identify features of radiatively efficient AGN activity (i.e. the HERGs) and separate these from radiatively inefficient AGN (i.e. the LERGs).

The recent Dark Energy Spectroscopic Instrument Early Data Release (DESI EDR; \citealt{DESICollaboration2023}) has provided spectroscopic observations for a moderate fraction ($\sim$ 20 per cent) of the LOFAR detected sources in the ELAIS-N1 and Bo\"{o}tes fields. To test the robustness of the source classifications using these new spectroscopic observations, we construct stacked (median) spectra for each of the HERG, SF-LERG, Q-LERG, and the SFG populations, as identified based on the classifications from \citet{Best2023}. The spectral stacking process used is presented and described in detail by \citet{Arnaudova2024}. In summary, we first apply Galactic extinction corrections using \citet{1998Schlegeldustmap} and the Milky Way reddening curve from \citet{fitzpatrick1999mwext} with $R_{\rm{V}} = 3.1$, and then de-redshift the individual spectra and resample them onto a common wavelength grid. These are then normalised at the reddest possible end where data from all spectra is available (this also minimises the effects of extinction). The stacked spectrum for each source type is then generated by computing the median in a given wavelength bin. The DESI selection function results in varying redshift distributions between each of the four samples; we hence limit the analysis to sources with spectroscopic redshifts in the range of $0.5 < z \leq 1$. To enable a more direct comparison to the results presented in the rest of the paper, we also only include sources with $L_{\rm{150\,MHz}} \geq 10^{24}\,\rm{W\,Hz^{-1}}$ when constructing the stacks.

The resulting stacked spectra for the HERGs, SF-LERGs, Q-LERGs, and SFGs are shown in Fig.~\ref{fig_ap:spec_stack}. Where applicable, we mark key emission line features in the spectra, namely, [O II]\,$\lambda 3728$, [Ne III]\,$\lambda 3870$, H$\beta$, and [O III]\,$\lambda \lambda 5007,4959$. The Q-LERG stack shows the Ca H and K absorption features, characteristic of the lack of young stars, along with a lack of emission line features that are typically indicative of star-formation or (radiatively-efficient) AGN activity; this is in-line with the expectations of Q-LERGs being the jet-mode (i.e. radiatively inefficient) AGN population. The HERG stack shows strong emission from the [O III] line (35.12\,eV), along with a weak detection of the higher ionisation [Ne III] line (40.96\,eV), both of which require the presence of a hard ionising radiation field from an AGN. The presence of other features such as H$\beta$ and the [O II] lines trace star-formation activity \citep[e.g.][]{Kennicutt1998} within the host galaxies of the HERGs. The strong [O III] emission seen for the HERGs, with a $\log_{10} \rm{([O III]/H\beta)} = 1.0$, gives further indication of AGN activity. If the SF-LERGs are indeed radiatively-inefficient AGN, like the Q-LERGs, but simply hosted in star-forming galaxies, we would expect the SF-LERG and SFG stack to present similar emission line features. Qualitatively, the SF-LERG stack is similar to the SFG stack, both showing stronger [O II] emission associated with star-formation compared to [O III]. The $\log_{10} \rm{([O III]/H\beta)}$ ratio for the SF-LERG stack is 0.7, which is lower than that of the HERGs, but higher than the value of 0.4 found for the SFG stack. Compared to the HERGs, the SF-LERGs also display a lower [O III]/[O II] ratio. Overall, these results suggest that while there may be a small level of contamination of SF-LERGs by low luminosity AGN, their average emission line features are broadly consistent with star-formation activity.

We note that the above analysis presents an average view of each source type and could only be performed at present for a subset of the radio sources with spectroscopic data available. The use of emission line ratio diagnostics, such as the BPT classification or excitation index, provides the most robust method of classifying LERGs and HERGs \citep[e.g.][]{2012MNRAS.421.1569B}. Such an analysis requires the detection and measurement of at least 4 -- 6 different emission lines, which are not always available for each source with DESI spectra, further reducing the number of sources for which such an analysis could be performed; we have therefore focused on performing spectral stacking in this paper. For the subset of radio sources detected within DESI EDR where emission lines can be measured robustly, such spectroscopic classifications will be presented in future work (Arnaudova et al. in prep.). Detailed spectroscopic classification for the vast majority of the radio sources will only be enabled by the upcoming WEAVE spectrograph \citep{Jin2024}, which will provide dedicated spectroscopic follow-up of LOFAR-detected sources, including \textit{all} LOFAR-Deep sources as part of the WEAVE-LOFAR survey \citep{2016sf2a.conf..271S}. This will overcome many of the limitations of existing SED-fitting based source classifications, including providing more reliable estimates of AGN bolometric luminosities and provide more robust redshifts. We note that for the shallower but wider LoTSS-DR2 \citep{Hardcastle2023}, where large SDSS spectroscopic samples are available, the classification of LERGs and HERGs is now possible using a radio-excess selection and BPT classification \citep{Drake2024}. This analysis, which focuses primarily at lower redshifts ($z \lesssim 0.5$) has also found the existence of a considerable population ($\sim$ 30 per cent) of LERGs hosted by star-forming galaxies. We expect this population to be less prevalent at lower redshifts but the fact that the use of the more robust emission line classifications still identifies a population of star-forming LERGs suggests that these are not simply due to mis-classifications from the SED fitting process.

\subsection{Average X-ray properties of LERGs and HERGs}\label{ap:xray_stack}
To further test the robustness of the source classifications, we investigate the X-ray properties of radio-detected AGN. X-ray observations can be used to trace radiatively efficient (radiative mode) AGN. For this analysis, we focus on the Bo\"{o}tes deep field which has coverage from deep \textit{Chandra} imaging over the full $\sim$9.6\,deg$^{2}$ of the field as part of the the \textit{Chandra} Deep Wide Field Survey (CDWFS;  \citealt{Masini2020}). The CDWFS covers 3.4\,Ms worth of imaging data across the soft (0.5 - 2\,keV), hard (2 - 7\,keV), and the broad (0.5 - 7\,keV) bands, reaching depths of $4.7\times 10^{-16}$, $1.5 \times 10^{-16}$, and $9 \times 10^{-16}$\,$\rm{erg\,cm^{-2}\,s^{-1}}$. In addition to the imaging datasets, \citet{Masini2020} also present a catalogue of 6891 X-ray detected sources across the three bands.

In addition to the LERGs (split into three sub-groups based on their star-formation activity) and the HERGs, we also investigate here the properties of the radio-quiet AGN (RQ-AGN), which are defined as sources that do not display a radio-excess but are identified as AGN based on SED fitting \citep[see][]{Best2023}. Like the HERGs, the RQ-AGN are thought to be powered by radiatively efficient AGN, and their comparison to the LERGs can provide additional insights. Throughout this analysis, our aim is to understand the properties of the AGN as identified and classified based on the radio data; our starting point is therefore the five classes of radio-detected AGN. We initially cross-match the radio-detected AGN with the \citet{Masini2020} X-ray catalogue (in the hard band), finding that around 33 per cent and 30 per cent of the HERGs and RQ-AGN, respectively, are also detected at X-ray wavelengths. In contrast, we find that only $\sim$ 2 per cent of the LERGs are X-ray detected (with the X-ray detections being split roughly equally amongst the three sub-groups). The RQ-AGN and the HERGs are both expected to be powered by a radiative-mode AGN which emit strongly in the X-rays due to inverse-Compton scattering of photons from the black hole accretion disc; the considerable fraction of both RQ-AGN and HERGs that are X-ray detected is therefore expected. LERGs on the other hand are expected to be powered by radiatively-inefficient accretion, which can explain the low X-ray detection rate. In addition, the similarly low X-ray detection fractions found for LERGs hosted within quiescent, intermediate, and star-forming galaxies indicates that the SF-LERGs are likely not mis-classified radiative-mode AGN. It is possible that some of the X-ray emission from the X-ray detected LERGs is associated with the radio jets \citep[e.g.][]{Hardcastle2009}.

\begin{figure}
    \centering
    \includegraphics[width=\columnwidth]{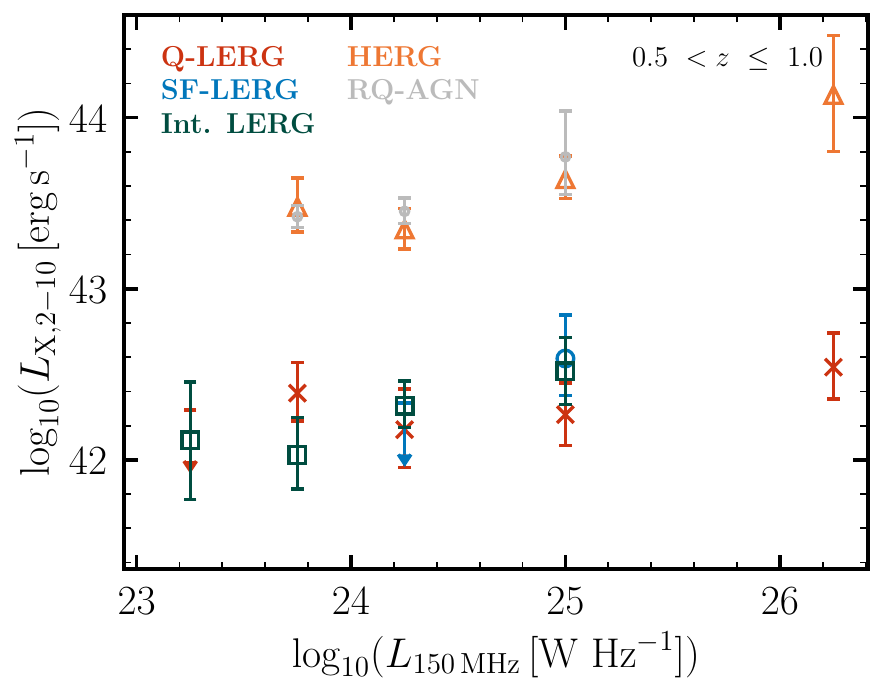}
    \caption{\label{fig_ap:xray_stack}The median stacked 2-10\,keV rest-frame X-ray luminosities as a function of radio luminosity for LERGs hosted by different galaxy types, HERGs, and RQ-AGN across $0.5 < z \leq 1$. The stacked X-ray luminosities represent the median X-ray luminosities based on bootstrap resampling, which are corrected for the contribution from XRBs. The uncertainties are computed using the corresponding bootstrap method. At a given radio luminosity, the LERGs, regardless of their host galaxy star-formation activity, show over an order of magnitude lower X-ray luminosities than the HERGs or RQ-AGN.}
\end{figure}

A significant fraction of our radio-detected AGN are however undetected in the X-rays; to gain a more complete understanding of the radio-detected AGN population we perform an X-ray stacking analysis to compute the average X-ray luminosities of each group of AGN; the full details of this process will be described by Holc et al. (in prep.). To ensure that there are sufficient sources in each group of AGN, we focus our analysis within the redshift range $0.5 < z \leq 1$. Over this redshift range, we split each AGN sample into five radio luminosity bins and derived the stacked average X-ray luminosities in each bin. Using the positions of the radio-detected sources, we calculated the stacked count rates and fluxes using the hard band (2-7\,keV) CDWFS images. We incorporated both the X-ray detections and non-detections into our stacking analysis as the detections constitute a significant fraction of the HERGs and RQ-AGN. To derive robust stacked X-ray luminosities and uncertainties, for each radio luminosity bin, we repeat this process 10,000 times by performing a bootstrap resampling of the input source list, to derive a distribution of 10,000 realisations of the average stacked X-ray fluxes. We took the median of this distribution as our best estimate of the stacked X-ray fluxes, with the corresponding 1$\sigma$ uncertainties estimated using the 16th and 84th percentiles. The stacked X-ray fluxes were then converted to rest-frame 2-10\,keV X-ray luminosities using the mean redshift of the bin ($z_{\rm{mean}}=0.75$) and assuming an X-ray power law spectrum with a photon index of $\Gamma = 1.8$.

The X-ray emission from a galaxy can arise not just from AGN activity but also from X-ray binaries (XRBs), which has been found to be well correlated with the star-formation activity \citep[e.g.][]{Lehmer2010,Mineo2014,Lehmer2016,Aird2017}. We estimate the contribution to the X-ray luminosity from XRBs using the empirical scaling relationship found by \citet{Lehmer2016} and subtract this from our stacked values. The resulting stacked 2-10\,keV rest-frame X-ray luminosity as a function of radio luminosity for different groups of AGN are shown in Fig.~\ref{fig_ap:xray_stack}. At a fixed radio luminosity, we find that both HERGs and RQ-AGN show significantly higher average X-ray luminosities, by over an order of magnitude, compared to the LERGs. Moreover, at a given radio luminosity, the stacked X-ray luminosities for LERGs hosted by galaxies of different star-formation activities are consistent with each other. These results suggest that LERGs within each of the quiescent, intermediate, and star-forming systems are not significantly contaminated on a population level by radiatively efficient AGN, which show much higher X-ray luminosities. The consistency between the HERGs and RQ-AGN provides further evidence that these are both powered by radiatively-efficient AGN. As X-rays trace the accretion activity onto the black hole, we would expect that these significant differences in the X-ray luminosities will also result in significant differences in the average black hole accretion rates between the LERGs and HERGs. Further detailed analysis of the average black hole accretion rates, including an investigation of their trends with radio luminosity and other galaxy properties will be investigated in future work (Holc et al. in prep.).


\bsp	
\label{lastpage}
\end{document}